\documentclass[reprint,superscriptaddress,aps,prl]{revtex4-2}

\usepackage[usenames]{color}
\usepackage{graphicx}
\usepackage{multirow}
\usepackage{verbatim}
\usepackage{amsmath}
\usepackage{amsfonts}
\usepackage{amssymb}
\usepackage{mathrsfs}
\usepackage{url}
\usepackage{color}      % use if color is used in text
\usepackage{float}
\usepackage{braket}
\usepackage{todonotes}
\usepackage[export]{adjustbox}
\usepackage{subfig}
\usepackage{setspace}
\usepackage{cleveref}
\usepackage{lipsum}

\usepackage{xr}
\makeatletter
\newcommand*{\addFileDependency}[1]{% argument=file name and extension
  \typeout{(#1)}
  \@addtofilelist{#1}
  \IfFileExists{#1}{}{\typeout{No file #1.}}
}
\makeatother

\newcommand*{\myexternaldocument}[1]{%
    \externaldocument{#1}%
    \addFileDependency{#1.tex}%
    \addFileDependency{#1.aux}%
}
\myexternaldocument{SI}

\begin{document}
\title{Analytical Theory of Near-Field Electrostatic Effects in Two-Dimensional Materials and van der Waals Heterojunctions}
	
\author{Qunfei Zhou}
\email{qunfei.zhou@northwestern.edu}
\affiliation{Materials Research Science and Engineering Center, Northwestern University, Evanston, IL 60208, USA}
\affiliation{Center for Nanoscale Materials, Argonne National Laboratory, Argonne, IL 60439, USA}
	
\author{Michele Kotiuga}
\email{michele.kotiuga@epfl.ch}
\affiliation{Theory and Simulation of Materials (THEOS) and National Centre for Computational Design and Discovery of Novel Materials (MARVEL), \'{E}cole Polytechnique F\'{e}d\'{e}rale de Lausanne, CH-1015 Lausanne, Switzerland}

\author{Pierre Darancet}
\email{pdarancet@anl.gov}
\affiliation{Center for Nanoscale Materials, Argonne National Laboratory, Argonne, IL 60439, USA}
\affiliation{Northwestern Argonne Institute of Science and Engineering, Evanston, IL 60208, USA}
	
\date{\today}

\begin{abstract}
We derive and validate a quantitative analytical model of the near-field electrostatic effects in the vicinity ($\gtrapprox 3$Å) of two-dimensional (2D) materials. In solving the Poisson equation of a near-planar point charge ansatz for the electronic density of a 2D material, our formula quantitatively captures the out-of-plane decay and the in-plane modulation of density functional theory (DFT)-calculated potentials. We provide a method for quickly constructing the electronic density ansatz, and apply it to the case of hexagonal monolayers (BN, AlN, GaN) and monochalcogenides (GeS, GeSe, GeTe, SnS, SnSe, SnTe, PbS, PbSe, PbTe) and their flexural and polar distortions. We demonstrate how our model can be straightforwardly applied to predict material-/angle-specific moir\'{e} potentials arising in twisted superlattices with periodicities beyond the reach of DFT calculations. 
\end{abstract}
	
\maketitle

Two-dimensional (2D) materials are hosts to a variety of desirable electrostatic and electrodynamics effects~\cite{Koppens2018Science, Louie2020NatComm,Javier2011NL,Agarwal2017PRB,Yakobson2017JACS,Lian2020PRL} that make them suitable for plasmonics and photonics applications~\cite{Avouris2014tHz,Ashwin2014NatPh,Wang2011GrtHz}.
In addition to these intrinsic properties, the properties of stacked and substrate-supported 2D  materials strongly depends on the near-field ($\approx 3$Å) electrostatic interactions between layers. For example, near-field electrostatic interactions in twisted bilayer hexagonal boron nitride (h-BN)~\cite{Pablo2021Science,Shalom2021FE,Woods2021BNFE,Wu2017FE} and transition metal dichalcogenides~\cite{Pablo2022FE,Weston2022FE,Wu2017FE,Wu2021FE} can induce macroscopic ferroelectric order, and control the excitonic energy landscape~\cite{Yao2015exciton,Xu2021excitons,Deng2020IX,Wurstbauer2017IX,Pasupathy2021moire}. With the discovery of exotic phenomena in moiré systems~\cite{Cao2018TBG,Dean2019TBG,Cao2021TTG,Cao2021TopGr}, accurate continuum models of the interlayer interactions and structural relaxations that connect macroscopic geometric descriptors to microscopic effective low-energy Hamiltonians have been proposed~\cite{Kaxiras2019TBG,Kaxiras2021Top,MacDonald2021PNAS,Li2021NatMatPhonon},  %MamorereferenceshereMayouMacDonald},
leading to plethora of predictions of exotic electronic states~\cite{MacDonald2018PRL,MacDonald2019TMD,Rubio2019NL,Sarma2020top,Rubio2021moire}. However, while the near-field electrostatic potential has been extracted from density functional theory (DFT)~\cite{Yao2021hBNV}, to the best of our knowledge, no simple closed form of the near-field interactions experienced in 2D materials assemblies or moiré structures has been proposed. 

In this work, we derive such an expression  for the electrostatic interactions and validate it extensively on DFT calculations. Our formula is obtained solving the Poisson equation for a near-planar discretized ansatz of the 2D materials charge density. We derive this ansatz for two families of 2D materials: hexagonal materials (h-AlN, h-BN, h-GaN) and transition metal monochalcogenides MX (M=Ge, Sn, Pb; X=S, Se, Te), and extract material descriptors for the strength and decay lengths of near-field interactions. At common interlayer distances ($3 \sim 4$ Å), we find that 2D materials can provide large in-plane ($0.1\sim 1$V/nm)  and out-of-plane ($0.1\sim 3$ V/nm) electric fields that both decay exponentially on the scale of the in-plane lattice vectors, effectively reducing the effect of nearest-layer interactions. Our formula also captures the effect of polar and finite-wavevector structural distortions, the latter of which exponentially impacts the electrostatic potential. Finally, we show how our model can easily predict angle-specific moir\'{e} potentials for large van der Waals heterostructures  with periodicities beyond the reach of DFT calculations.

\begin{figure}[h]
	\centering
	\includegraphics[width=0.95\linewidth, center]{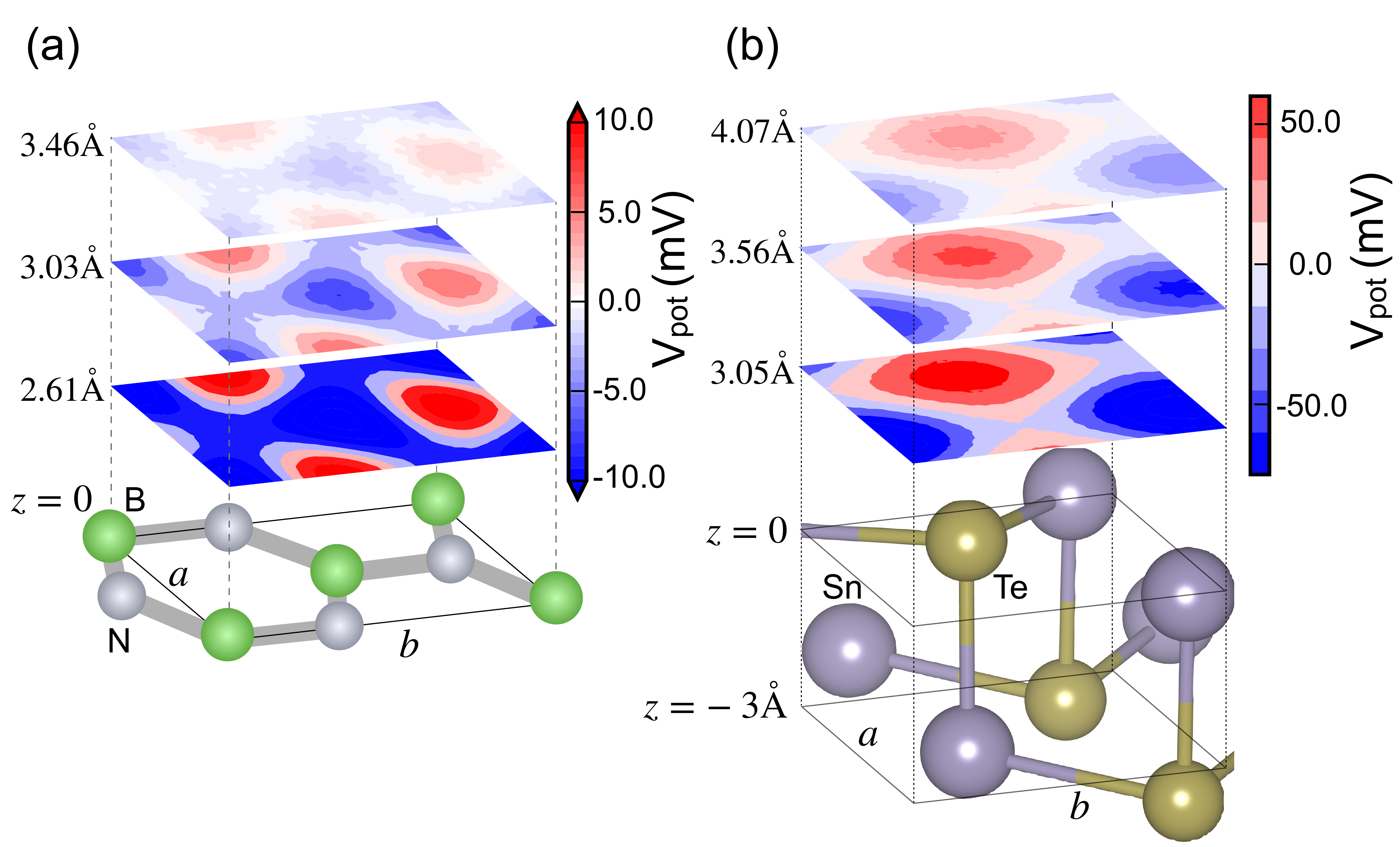}
	\caption{In-plane electrostatic potential above a single layer of 2D (a) h-BN and (b) SnTe, as a function of out-of-plane distance computed with DFT. The potential values are referenced to that of vacuum (V$_{\textrm{vacuum}}$=0V).} 
	\label{fig:f1}
\end{figure}

As shown in  Fig.~\ref{fig:f1}, 2D material monolayers, such as h-BN and SnTe, induce a modulation of electrostatic potentials at typical 2D-2D and 2D-substrate distances in 2D van der Waals heterostructures ($\simeq 3.3$\AA\ and $\simeq 3.1$\AA\ for bilayers of h-BN and SnTe, respectively). All DFT parameters and computational details are included in the supplementary information. The potential increases significantly at smaller distances, and decays to negligible values at larger distances ($d=$ 5\AA) for both 2D materials. In the following and as shown in Fig.~\ref{fig:f1}, we define the coordinates of a point $(x,y,d)$ with respect to the $z=0$ plane computed from the averaged atomic coordinates in the top layer (e.g boron and nitrogen are located in the $z=0$ plane for h-BN, and Sn and Te dimers are located near the $z=0$ and $z=-3.0$Å planes for SnTe). All atomic structures are included in supplementary information. 

\begin{figure}[h]
	\centering
	\includegraphics[width=0.99\linewidth, center]{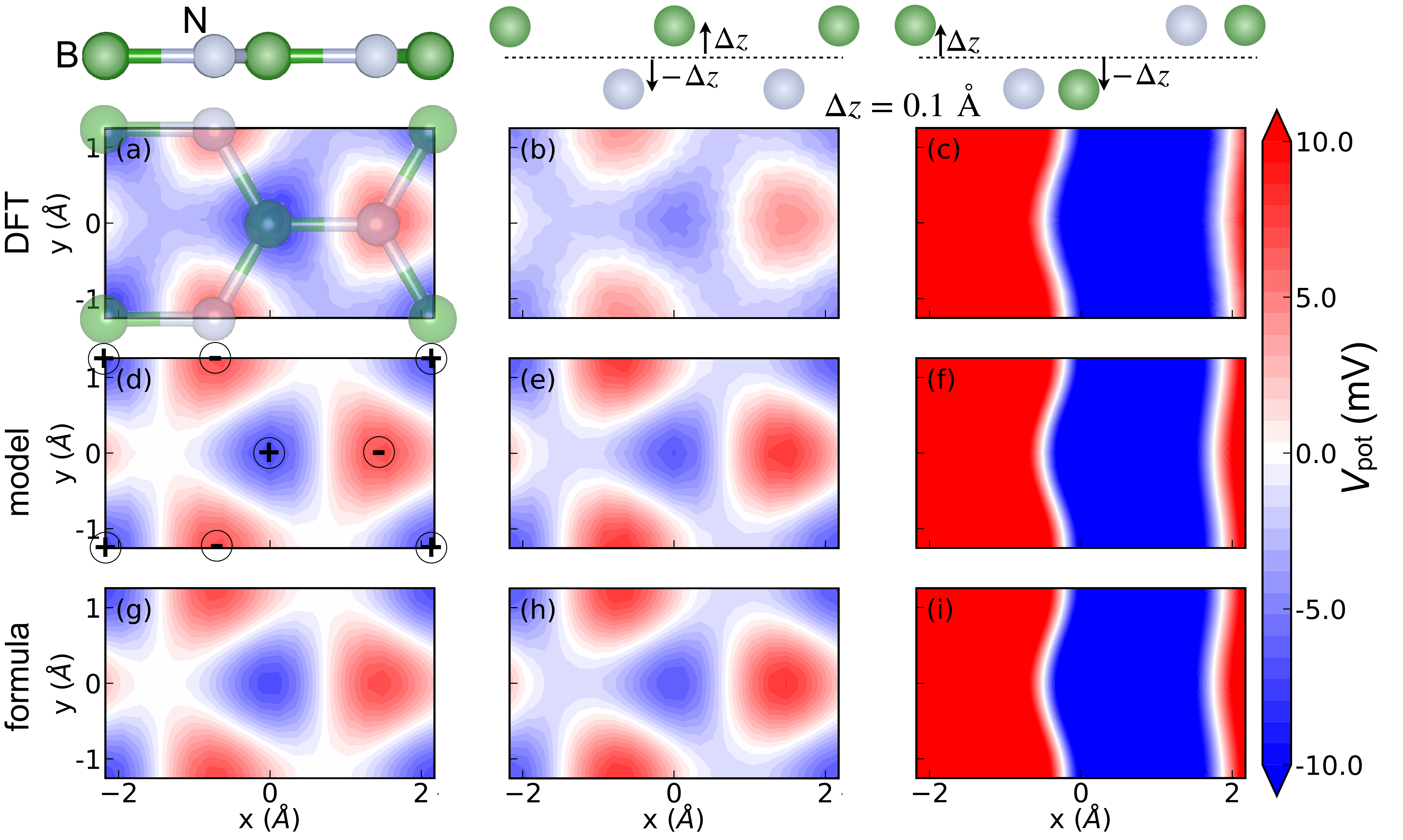}
	\caption{In-plane electrostatic potential at $d=3.03$ \AA\ from single layer of h-BN computed from (a-c) DFT, (d-f) the DCD model, (g) Eq.~\ref{eq:BN}, (h) Eq. S13, (i) Eq. S15. The columns correspond (from left to right) to the potential above the pristine h-BN structure, h-BN deformed along the ZO infrared-active, and h-BN deformed along the ZA mode at the corner of the Brillouin zone (see corresponding atomic configurations on the top).} 
	\label{fig:BNz}
\end{figure}

The magnitude and shape of this in-plane modulation, as well as the lengthscale of its far-field decay, are material-dependent and retain some information of the underlying atomic structure. For example, in the $d=3$Å  plane, the $\simeq14$ meV in-plane modulation of the electrostatic potential of h-BN has minima (i.e. is most attractive to electrons) close to the positively-charged borons and maxima near the negatively-charged nitrogens, as shown in Fig.~\ref{fig:BNz}(a,d). The location of these maxima and minima is weakly impacted by the vertical distance and the exponential~\cite{Natan2007SAMrev} decay in the far-field with  an out-of-plane decay constant $\beta =2.80$\AA$^{-1}$.

\begin{figure}[h]
	\centering
	\includegraphics[width=0.95\linewidth, center]{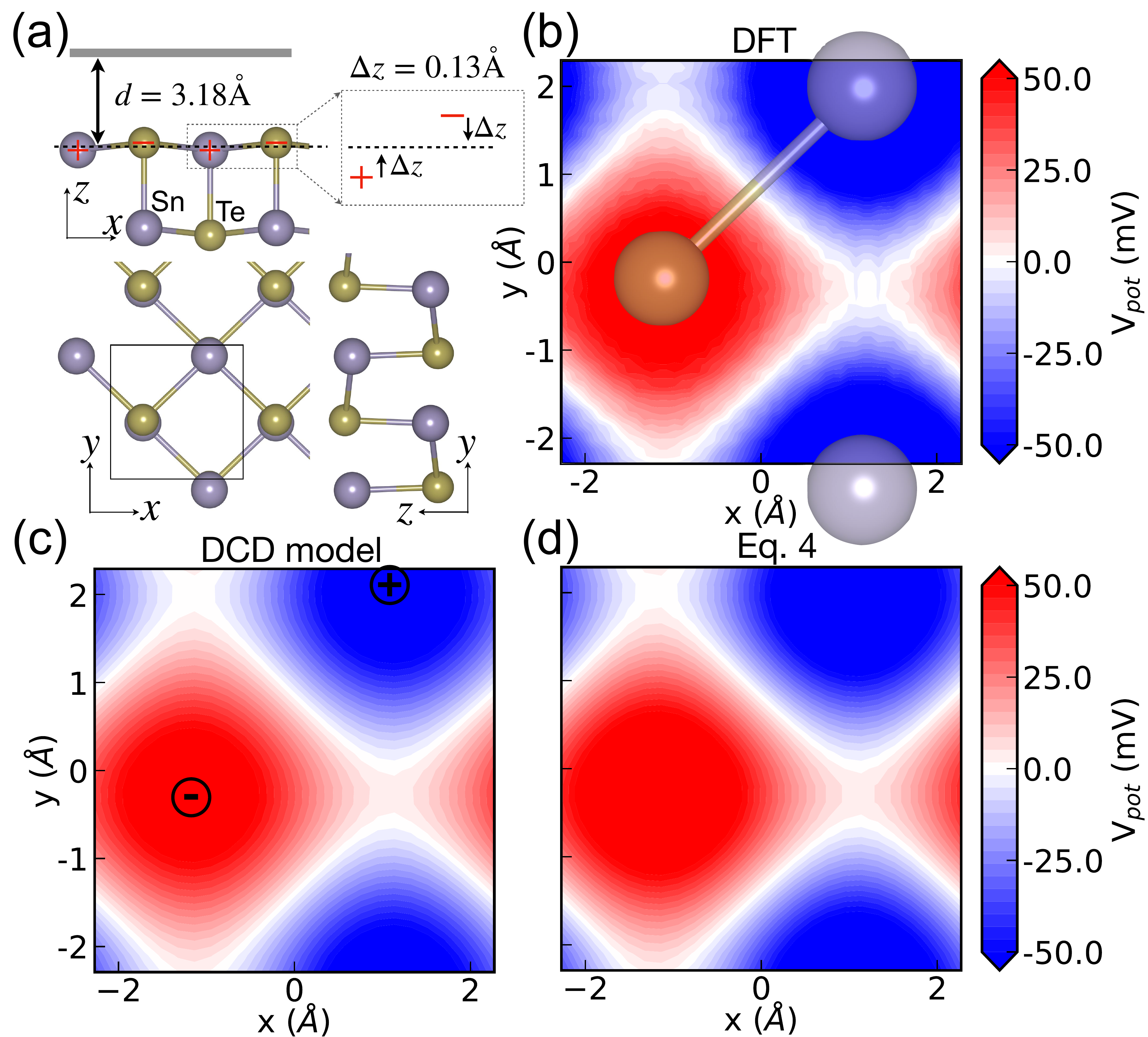}
	\caption{(a) Structure of SnTe, and in-plane electrostatic potential at $d=3.18$ \AA\ from the plane of single layer SnTe computed from (b) DFT, (c) DCD model, and (d) Eq.~\ref{eq:mmc2}.} 
	\label{fig:SnTe1d}
\end{figure}

At similar vertical distance of $\sim$3.0\AA, the in-plane modulation of the  potential is an order of magnitude larger (150 meV) in SnTe, with the position of the extrema close to the top-most atoms in the monolayer, see Fig.~\ref{fig:SnTe1d}(b,c). The out-of-plane decay constant is also much weaker ($\beta =1.19$\AA$^{-1}$), in accordance with the larger in-plane lattice parameters~\cite{Natan2007SAMrev,Woods2021BNFE,Yao2021hBNV, Zhou2021SAM}. This is in agreement with previous works~\cite{Natan2007SAMrev, Yao2021hBNV, Zhou2021SAM} and suggests that the electrostatic potential at such distances can be well approximated by a functional form $V(x,y,d) \simeq f(x,y) e^{-\beta d} $, with $f(x,y)$ depending on the atomic positions and $\beta$ depending on the lattice parameter and symmetry, which we now derive. 

Our derivation consists of two steps: (1) approximate the continuous electronic density of the 2D layer by a set of discrete point charges near the atomic positions, i.e. $\rho(x,y,z) = \sum_i q_i \delta(\mathbf{r} - \mathbf{r}_i)$, and (2) solve the Poisson equation analytically assuming that the vertical fluctuation of these point charges from the plane, $\Delta z_i$, is small compared to $d$, and that $\sum_i q_i =0$. The conditions,  $\Delta z_i<< d$ and $\sum_i q_i =0$, imply that the monochalcogenide compounds and/or multi-layer 2D materials are approximated by a sum of several neutral slices (e.g. two neutral slices near z=0 and z$\simeq$-3.0Å for SnTe). As discussed more in the next paragraphs and as shown in Fig.~\ref{fig:f1}, the effect of the atoms in the bottom plane of monochalcogenides and/or of multilayer substrates is exponentially weaker and can easily be neglected.      

Inspired by atomic representations of electrostatic effects in molecular dynamic simulations~\cite{Truhlar2014Charge} and with the objective to reproduce the DFT-calculated electrostatic potential, we determine $\{q_i,\mathbf{r}_i\}$ following the procedure outlined in  our previous work on self-assembled molecular layers~\cite{Zhou2021SAM}. In contrast to~\cite{Zhou2021SAM}, (i) we set the number of point charges to the number of atoms in a given neutral slice (e.g. 2 point charges per unit cell in h-BN, 2 slices of 2 point charges per unit cell in monochalcogenides), (ii) we allow the point charges in a given slice to be non-planar, as shown in Fig.~\ref{fig:SnTe1d}(a), %\textcolor{red}{but restrict 
and assigning their $(x,y)$ position to be at the position of the potential extrema, unlike the model of atomic charges~\cite{Truhlar2014Charge}. We refer to this model as discretized charge density (DCD) model. The resulting DCDs for h-BN, h-AlN, h-GaN,  and all MX (M=Ge, Sn, Pb; X=S, Se, Te) are reported in Figs.~\ref{fig:SnTe1d}(c), ~\ref{fig:BNz}(d) and the supplementary information. As shown in Fig. S6, the constraint of the in-plane positions of the point charges means they are found close to, but not necessarily at, the atomic locations. In the case of SnTe, as shown in Fig.~\ref{fig:SnTe1d}(a-c), the positive and negative point charges are located 0.07\AA\ and 0.16\AA\ away from the Sn and Te atoms, respectively.
%\textcolor{red}{???}Å away from the atoms ($\Delta z \simeq \pm 0.1$\AA\ for Te and Sn, respectively), reminiscent of the fact the extra charge is carried by the Te lone pair. 

In Fig. \ref{fig:SnTe1d} (b,c), we show the results from a 2-charge DCD model of SnTe when compared with DFT results of the same structure. As shown in Fig. \ref{fig:SnTe1d} (c) and in supplementary information for other monochalcogenides, the magnitude of the potential fluctuations and their locations are quantitatively captured by the DCD model, with a mean square deviation on the scale of 0.1meV  at d=3.18Å, 0.02 meV at d=4.0Å, 0.01 meV at d=5.0Å, and a mean square deviation of 0.02\AA$^{-1}$ on the decay constant. For reasons elaborated upon below, only charges in the top plane of SnTe are necessary to obtain a quantitative agreement. Results for h-AlN, h-GaN and all monochalcogenides are included in the supplementary information. 

With this DCD representation, we can write the potential as the interaction with the point charges and their periodic replica, e.g. (for an orthorhombic lattice of lattice constants a,b) as: 
\begin{align}
    V(x,y,z) & = k_e \sum_\textrm{plane} \sum_{n_1,n_2} \sum_{i\in\textrm{plane}}  q_i \left[ (x-x_i -n_1 a)^2 \right. \nonumber \\
    & \left. + (y-y_i - n_2 b )^2 + (z-z_\textrm{plane} - \Delta z_i )^2\right] ^{-1/2}
\end{align}
where $n_1$ and $n_2$ are integers to sum over the (infinite number of) periodic images, and $k_e$ is the Coulomb constant.

As shown in the supplementary information and in agreement with our previous work~\cite{Zhou2021SAM}, this particular sum converges much faster in reciprocal space. Specifically, the analytical form in reciprocal space for a given neutral plane at a distance $d$ is

\begin{align}
& V(x,y,d)   = k_e\sum_i q_i \left[\frac{2\pi \Delta z_i}{ab} + \right. \nonumber \\
& \qquad \left. \sum_{\vec{k}}{}' 
\frac{e^{-2\pi\left[ \frac{i k_1(x-x_i)}{a}+\frac{i  k_2(y-y_i)}{b}+(d-\Delta z_i) \sqrt{\frac{k_1^2}{a^2}+\frac{k_2^2}{b^2}}\right]}}
{\sqrt{b^2k_1^2+a^2 k_2^2}}\right].  \label{eq:final}
\end{align}
 The first term in the sum corresponds to the areal dipole term identified by Natan et al.\cite{Natan2007SAMrev}.  The second sum of $\vec{k}=[k_1,k_2]$ runs over the reciprocal lattice points excluding the origin. The lattice parameters $a,b$ and absolute magnitude of $q$ are summarized in Table S1 for h-BN, h-AlN, h-GaN and all monochalcogenides.

Eq.~\ref{eq:final} is the %central result 
central result of this work. Unlike its real space expression, this expression converges with a few $\vec{k}=[k_1,k_2]$, allowing us to derive an analytical solution for the potentials from 2D materials. For h-BN, h-AlN, and h-GaN, only 4 terms are necessary to obtain a $\geq$99\% estimate of the full sum. In the case of monochalcogenides, this value is decreased to 3 terms. 
 
Specifically, writing Eq.~\ref{eq:final} for monolayer h-BN, h-AlN, and h-GaN, we obtain, 
\begin{align}\label{eq:BN}
V(x,y,d) & =k_e\frac{2A(x,y)q}{a} e^{\frac{-4\pi d}{\sqrt{3}a}}
\end{align}
where $A(x,y)=-2\sqrt{3}\sin(2\pi y-\frac{\pi}{3})\left[\cos(2\pi x)+\cos(2\pi y-\frac{\pi}{3})\right]$. $x, y$ are in fractional coordinates of the atomic positions. $a$ is 2.51\AA, 3.13\AA, and 3.21\AA\ for h-BN, h-AlN, and h-GaN, respectively. We note that the out-of-plane $e^{\frac{-4\pi d}{\sqrt{3}a}}$ decay for h-BN has been previously reported in the literature~\cite{Yao2021hBNV}. 

Similarly, for monochalcogenides, Eq.~\ref{eq:final} can be simplified to:
\begin{align} \label{eq:mmc2}
& V(x,y,d) = k_e \frac{q}{a}  \left\{ B_1(x,y,\Delta z) e^{-2\pi d\sqrt{1/a^2+1/b^2}} \nonumber \right. \\
 & \qquad \quad \left. + B_2(x,y,\Delta z) e^{\frac{-2\pi d}{b}} + B_3(x,y,\Delta z) e^{\frac{-2\pi d}{a}}  \right\}
\end{align}
where $B_1(x,y,\Delta z)$, $B_2(x,y,\Delta z)$, $B_3(x,y,\Delta z)$ are trigonometric functions of ($x,y,\Delta z$), given in supplementary information. The rapid out-of-plane $e^{\frac{-2\pi d}{a}}$ decay, also explains that only the charges at the top plane contribute appreciably, as this factor decreases the impact of the bottom plane by $\geq 95\%$ at d$\simeq$3.18 Å, e.g., the total potential modulation for SnTe at d$\simeq$3.18 Å is $\simeq$154 meV while the bottom layer only contribute 2.79 meV. 

As shown in Fig. \ref{fig:SnTe1d} (d) and in the supplementary information, despite their simplicity, Eq.~\ref{eq:BN}-Eq.~\ref{eq:mmc2} are in quantitative agreement with DFT results at distances $d > 2.5$Å. 

Moreover, Eq.~\ref{eq:BN}–Eq.~\ref{eq:mmc2} have significant implications on the wavelength dependence of structural distortions on near-field interactions. In particular, while the near-field electrostatic effects of the 2D materials decays exponentially with normal distance $d$ on a length scale fixed by lattice parameters of the unit cell, the periodic corrugation of the atoms, $\Delta z$, only appears in the trigonometric forms $A(x,y,\Delta z)$ and $B_i(x,y,\Delta z)$. This implies that periodic corrugation effects, such as infrared(IR)-active phonons at $q=0$, will only cause a linear change in the near-field potential, while finite wavevector modes will have an exponential effect.

To test this hypothesis, we consider the impact of two different out-of-plane structural distortion modes on the near-field electrostatic potential of h-BN shown in Fig.~\ref{fig:BNz}, one corresponding to the q=0 ZO polar mode stabilizing the ferroelectric order in~\cite{Pablo2021Science,Shalom2021FE,Woods2021BNFE,Wu2017FE}, and the other corresponding to a ZA mode at the corner of the Brillouin zone. For illustrative purpose, we consider the same amplitude of displacement ($\Delta z =0.1$ Å, in the order of out-of-plane atomic displacement in bilayer h-BN at domains of different stackings\cite{Shalom2021FE}) %\textcolor{red}{taken from \cite{Shalom2021FE}?}) 
while noting that the amplitude along the ZA mode would be larger under thermal occupation. In the right two columns of Fig.~\ref{fig:BNz}, we show the atomic structures with out-of-plane displacement of $\Delta z=0.1$\AA\ and potentials at $d\simeq 3.0$ from the averaged position of the atoms (the black dashed lines). In both cases, the DCD model and analytical formulas (Eq.~\ref{eq:BN}, Eq. S13, Eq. S15) correctly predict the large fluctuation of the near-field potential in response to the finite wavevector (Fig.~\ref{fig:BNz} (e-f), (h-i)), as compared with DFT results (Fig.~\ref{fig:BNz} (b-c)). Fig.~\ref{fig:BNz} also implies that our model can also be used to explore the electrostatic impacts of strain and corrugation in the properties of moiré structures~\cite{Crommie2021moire,Shalom2021FE}.

Fig.~\ref{fig:BNz} further emphasizes the dominant role of the in-plane periodicity in describing the near-field electrostatic potential~\cite{Natan2007SAMrev}: the doubling of periodicity in Fig.~\ref{fig:BNz} (c) results in a strong modulation on the shape and the magnitude of the potential fluctuation. When compared to Fig.~\ref{fig:BNz} (b), at equal amplitude of distortion and at $d\simeq3.0$Å, the modulation is about five times larger (potential fluctuation is 14 meV in Fig.~\ref{fig:BNz} (b) vs. 68 meV in Fig.~\ref{fig:BNz} (c)), and the potential becomes quasi-one-dimensional.  We also note that, while the magnitude of the point charge $q$ impacts the potential, the effect is also only linear and can be neglected as compared to the effects of lattice parameters (SnTe has much larger potential modulation than h-BN, despite $q=  0.18$ for SnTe and $q=0.89$ for h-BN, see Table S1). 

With this framework, we finally demonstrate how our analytical formulas can be used to predict moir\'{e} potential superlattices for twisted van der Waals heterostructures, which are challenging for DFT calculations due to the large number of atoms in the unit cell. We note that we are neglecting the structural relaxations that robust models have been already proposed and validated~\cite{Kaxiras2019TBG,Kaxiras2021Top}, and focus solely on the electrostatic component in the following. Under these assumptions, the moiré potential can be obtained by a simple rotation of the atomic coordinates in one of the layers, i.e. by solely modifying the trigonometric forms $A(x,y,\Delta z)$ and $B_i(x,y,\Delta z)$ in Eq.~\ref{eq:BN} and Eq.~\ref{eq:mmc2} for hexagonal monolayers and monochalcogenides, respectively. 

\begin{figure}[h]
	\centering
	\includegraphics[width=0.99\linewidth, center]{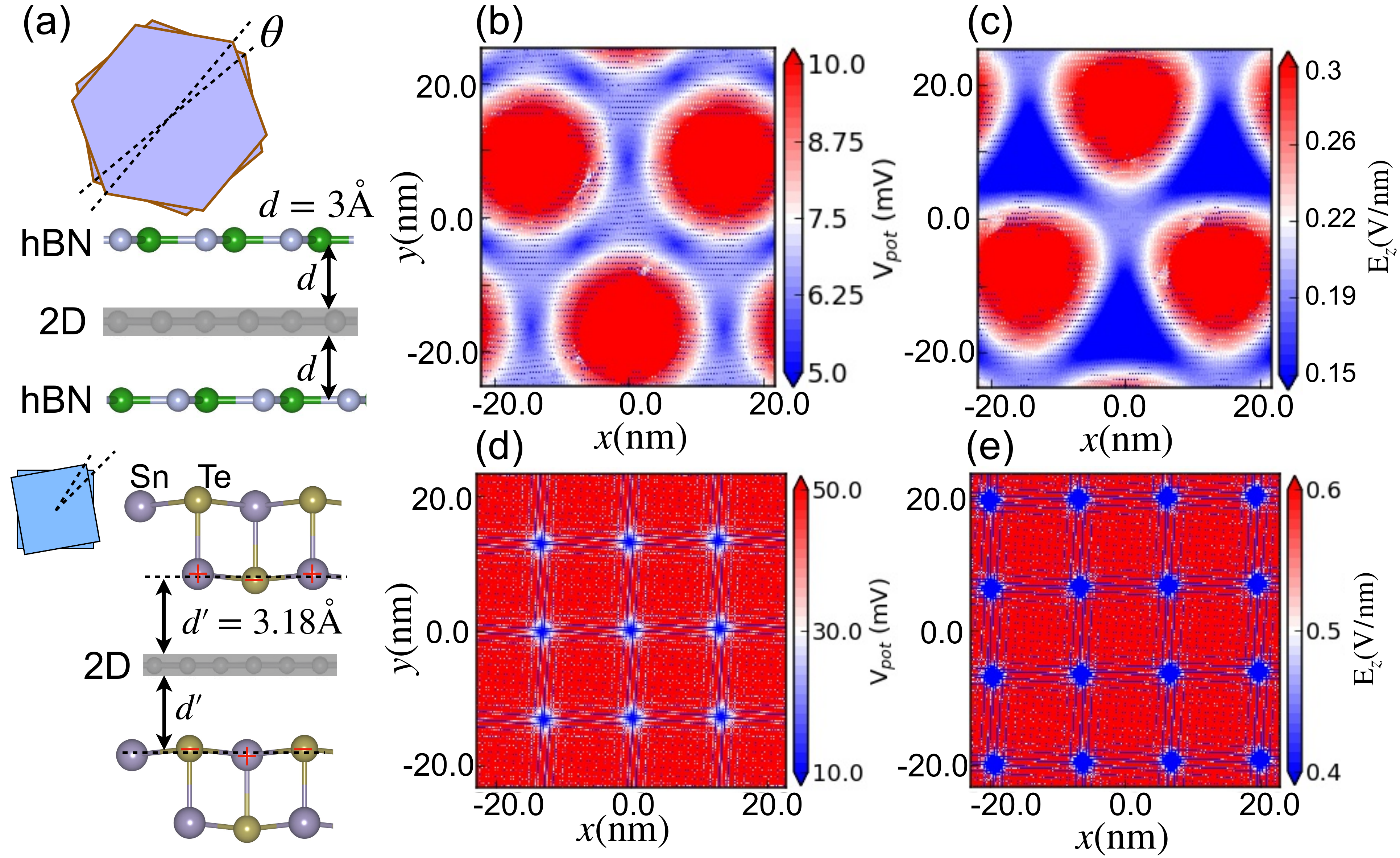}
	\caption{(a) A schematic showing the twist in rotation with angle $\theta$ of two layers of h-BN (SnTe), and a sandwich structure with two h-BN (SnTe) layer at the top and bottom of another 2D layer. (b) Electrostatic potential and (c) out-of-plane electric field at the position of the center 2D layer when the two h-BN layers are twisted by 0.5$^o$. Same for SnTe with a twist angle of 2$^\textrm{o}$ in (d) and (e). %The moir\'{e} wavelength $\lambda$ is $\sim$28 nm and $\sim$19 nm for h-BN and SnTe, respectively.
	} 
	\label{fig:moire}
\end{figure}

We adopt a heterostructure configuration that has demonstrated exotic property control by the twist angle of the top and bottom layer\cite{Dean2018BNGrBN,Moire2021BNGrBN}, as shown in Fig.~\ref{fig:moire}(a) and compute the potential experienced by a  central planar 2D material, surrounded by top and bottom h-BN or SnTe. When the two h-BN layers are twisted by angle $\theta$, the potential becomes:
\begin{align}
    V(x,y,d)=k_e\frac{2qA^\prime(x,y,\theta)}{a}e^{\frac{-4\pi d}{\sqrt{3}a}} \label{eq:moireBN}
\end{align}
where $A^\prime(x,y,\theta)=A(x,y)+A(x^\prime,y^\prime)$ with $A(x,y)$ same as that in Eq.\ref{eq:BN} and $x^\prime= x\cos \theta + by\sin \theta/a, y^\prime=-ax\sin \theta/b + y\cos \theta$. This also allows us to compute the resulting electric field (Eq. S24). In Fig.~\ref{fig:moire} we present the in-plane potential and out-of-plane electric field at the center 2D layer for h-BN with rotation angle $\theta=0.5^\textrm{o}$ and SnTe with $\theta=2^\textrm{o}$. The twisted bilayer h-BN generates a moir\'{e} potential superlattices with wavelength on the order of 30nm at $\theta=0.5^o$, a lengthscale challenging for DFT calculations. As 2D materials can be controlled and manipulated via potential modulations or external electric fields\cite{Park2008NatPhys,Park2008PRL,Li2021Gr, Ma2021ExcitonTraps,Zhou2021SAM}, our simple analytical model for near-field electrostatic effects of layered 2D materials can be used for study the electronic and optical response of various moir\'{e} potential superlattices.
%electric field, positive, out-of-plane polarization

In summary, we have derived a novel material-specific analytical theory of the large near-field electrostatic effects in 2D materials. Our theory fully captures the magnitude of the potential modulations (on the order of $\simeq10$ mV for h-BN and $\simeq100$mV for monochalcogenides) at typical interlayer  distances (3\AA\ $\sim$ 4\AA) of van der Waals layered structures, as well as the out-of-plane decay of these fluctuations. We parametrized a discretized charge density model to reproduce DFT-calculated potentials for two classes of 2D materials. Importantly, our formula is predictive and elucidates the importance of in-plane lattice constants, geometric effects, structural distortions, corrugation, distance, and material polarity in the electrostatic potential. Furthermore, our formula can be used to compute an analytical expression of the angle-specific moiré potentials  and electric fields on lengthscales on the order of tens to hundreds of nanometers, which are challenging for DFT calculations. 

\section*{Supplementary Material}
Details of the computational methods, including the details for DFT calculations, and the DCD model; Results of the electrostatic potential for h-AlN, h-GaN and all monochalcogenides from DFT, DCD model and analytical formula; Potential fluctuations as a function of distance for all 2D materials from DFT, DCD model and analytical formula;
Structures of the 2D materials in crystallographic information file (cif) format. \\

\section*{Acknowledgments}
This work was supported by the Northwestern University MRSEC under National Science Foundation grant No.~DMR-1720139 (Q.Z., P.D.). Work performed at the Center for Nanoscale Materials, a U.S. Department of Energy Office of Science User Facility, was supported by the U.S. DOE, Office of Basic Energy Sciences, under Contract No. DE-AC02-06CH11357. We gratefully acknowledge use of the Bebop cluster in the Laboratory Computing Resource Center at Argonne National Laboratory. P.D. would like to acknowledge fruitful discussions with Cristian Cortès, Lincoln Lauhon, and Mark Hersam.

\bibliography{Biblio/MultipoleDoping.bib}

\end{document}

\title{Supplementary Information for Analytical Theory of Near-Field Electrostatic Effects in Two-Dimensional Materials and van der Waals Heterojunctions}
	
\author{Qunfei Zhou}
%\email{qunfei.zhou@northwestern.edu}
\affiliation{Materials Research Science and Engineering Center, Northwestern University, Evanston, IL 60208, USA}
\affiliation{Center for Nanoscale Materials, Argonne National Laboratory, Argonne, IL 60439, USA}

\author{Michele Kotiuga}
\email{michele.kotiuga@epfl.ch}
\affiliation{Theory and Simulation of Materials (THEOS) and National Centre for Computational Design and Discovery of Novel Materials (MARVEL), \'{E}cole Polytechnique F\'{e}d\'{e}rale de Lausanne, CH-1015 Lausanne, Switzerland}
	
\author{Pierre Darancet}
\email{pdarancet@anl.gov}
\affiliation{Center for Nanoscale Materials, Argonne National Laboratory, Argonne, IL 60439, USA}
\affiliation{Northwestern Argonne Institute of Science and Engineering, Evanston, IL 60208, USA}
	
\date{\today}

\maketitle

\tableofcontents

\section{Computational Methods}	\label{SecSI:method}
%\subsection{Computational Details} \label{Sec:DFT}
All first-principles calculations based on Density Functional Theory are performed using the Quantum-Espresso package~\cite{Giannozzi2009QE,Giannozzi2017QE}, with Optimized norm-conserving Vanderbilt pseudopotentials~\cite{Hamann2013} obtained from the PseudoDojo library~\cite{Van2018pseudodojo}. Exchange-correlation potentials use Perdew-Burke-Ernzerhof (PBE) parametrization of the generalized gradient approximation~\cite{Perdew1996}. The plane-wave cutoff energy is 90 Ry for all systems studied in this work. $k$-point sampling of 16$\times$16 is used for h-BN with in-plane size of 2.51$\times$2.51 \AA, or equivalent for other cell sizes. The vacuum regions are all larger than 21 \AA. Convergences of total and electronic energies are 10$^{-5}$ eV/atom, and $10^{-6}$ eV, respectively. 

\section{Analytical Derivation: Periodic 2D array of charges not contained within a plane}

The potential from a periodic charge configuration, where the unit cell is a rectangular lattice with lattice constants of $a, b$ is:
\begin{align}
V(x,y,z=d) = k\sum_{n_1,n_2,i} \frac{q_i}{\sqrt{(n_1a+x_i-x)^2+(n_2b+y_i-y)^2+d^2}} \quad \& \quad \sum_i q_i = 0 \label{eq:1}.
\end{align}

Using Fourier transform, see detailed derivation in our previous work\cite{Zhou2021SAM}, Eq.\ref{eq:1} becomes:
\begin{align} \label{eq0}
	V(x,y,z=d)
	%&= k_e \sum_i\frac{q_i}{a} \zeta\left(\frac{1}{2}\right)\\   
	=k_e \sum_i \frac{q_i}{a} \sum_{\vec{k}}{}'
	\frac{e^{-2\pi\left[ i k_1(x-x_i)/a+ i  k_2(y-y_i)/b+d \sqrt{k_1^2/a^2+k_2^2/b^2}\right]}}
	{\sqrt{k_1^2b^2/a^2+k_2^2}} 
\end{align}

We can extend this formulation to consider a neutral periodic 2D array of charges that are not contained within one plane, but have varying distances along the perpendicular direction. In this case Eq.~\ref{eq:1} becomes 
\begin{align}
V(x,y,z=d) = k_e\sum_{n_1,n_2,i} \frac{q_i}{\sqrt{(n_1a+x_i-x)^2+(n_2b+y_i-y)^2+(z_i-d)^2}} \quad \& \quad \sum_i q_i = 0 \label{eq:2}.
\end{align}
$\delta$ in Eq.S12 in the SI of our previous work\cite{Zhou2021SAM}  is now defined as $\delta = (d-z_i)/a$, which leads to a modified form of the final expression, eq.~\ref{eq:final}, as 
\begin{align}
V(x,y,z=d)  
&=k_e\sum_i \frac{q_i}{a} \left[\frac{2\pi z_i}{b} + \sum_{\vec{k}}{}' 
\frac{e^{-2\pi\left[ i k_1(x-x_i)/a+ i  k_2(y-y_i)/b+(d-z_i) \sqrt{k_1^2/a^2+k_2^2/b^2}\right]}}
{\sqrt{k_1^2b^2/a^2+k_2^2}}\right]. \label{eq:final}
\end{align}
We note that in the case where the charge distribution yields a net dipole the first term is the potential  due to a periodic array of dipoles in the far-field limit, which can be equated to an infinite parallel-plate capacitor. Note in order to recover the expected potential difference, we need to calculate the potential different between $z=\pm d$, which picks up a factor of two. %The potential of a periodic array of dipoles will be considered in the next section. 

\subsection{Formula Derivation for hexagonal BN, AlN and GaN}
\begin{figure}[H]
	\centering
	\includegraphics[width=0.35\linewidth]{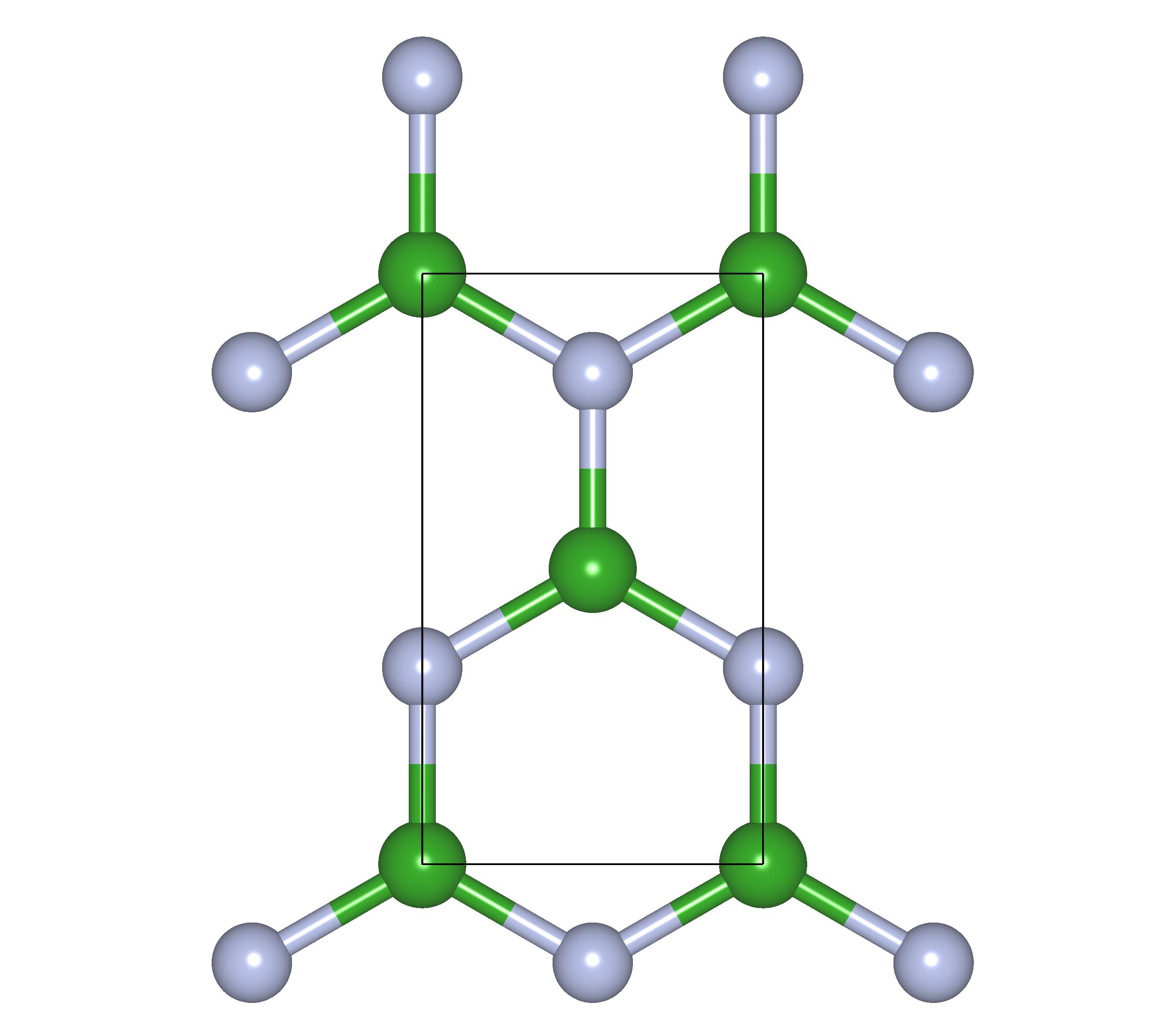}
	\caption{An Orthorombic Cell of monolayer h-BN, same for AlN and GaN where the N atoms are in gray.} \label{fig:BNcell}
\end{figure}
As shown in Fig.~\ref{fig:BNcell}, each orthorombic cell of h-BN (same for 2D AlN and GaN) with lattice parameters $a, b$, has four atoms (two B atoms and two N atoms), two B atoms with charge $q$ at position $\bs r(x,y,z=0)$ of  $(0,0,0), (-1/2a,-1/2b,0)$ and two N atoms with charge $-q$ at $(0,1/3b,0), (-1/2a, -1/6b, 0)$ when defining the h-BN plane at $z=0$. The potential at position $(x,y,z=d)$ is then (\emph{hereinafter, $x,y$ are in fractional coordinates, and $d$ has the same unit as $a$ and $b$}):
\begin{align}
	V(x,y,z=d) & = k_e \frac{q}{ab} \sum_{\vec{k}}{}' \frac{e^{-2\pi d M_k}}{M_k} \left\{ \cos[2\pi (k_1x+k_2y)] +  \cos[2\pi(k_1(x-\frac{1}{2})+k_2(y-\frac{1}{2})]\right. \nonumber \\ 
	& \left. {\qquad} - \cos[2\pi(k_1x+k_2(y-\frac{1}{3}))] + \cos[2\pi k_1(x-\frac{1}{2})+2\pi k_2(y-\frac{5}{6})] \right\} \\
	& = k_e\frac{4q}{ab} \sum_{\vec{k}}^{k_1+k_2=2n}{}' \frac{e^{-2\pi d \sqrt{k_1^2/a^2+k_2^2/b^2}}}{\sqrt{k_1^2/a^2+k_2^2/b^2}} \left[- \sin(2\pi k_1x + 2\pi k_2y -\frac{1}{3}\pi k_2)\sin(\frac{\pi k_2}{3})\right]
\end{align}

We could achieve $>$ 99\% accuracy with $k_1=[-1,1], k_2=[-2,2]$, and $b=\sqrt{3}a$. Therefore,
\begin{align}\label{eq:BN}
%		V(x,y,d) & =k_e\frac{4\sqrt{3}q}{a} \sin(2\pi y-\frac{\pi}{3})\left[ \cos(2\pi x) + \cos(2\pi y-\frac{\pi}{3})\right] e^{\frac{-4\pi d}{\sqrt{3}a}}
		V(x,y,d) & =k_e\frac{2A_1q}{a} e^{\frac{-4\pi d}{\sqrt{3}a}}
\end{align}
where $A_1=-2\sqrt{3}\sin(2\pi y-\frac{\pi}{3})\left[\cos(2\pi x)+\cos(2\pi y-\frac{\pi}{3})\right]$

The in-plane potential modulation defined as $\Delta V$ is:
\begin{align}
	\Delta V & = V_{max} - V_{min} = V(0,1/3b,d) - V(0,0,d)  \nonumber \\
	& = k_e\frac{18q}{a} e^{\frac{-4\pi d}{\sqrt{3}a}}
\end{align}

In the case of a rotation of the h-BN plane in angle $\theta$, the potential keeps the same form as Eq.~\ref{eq:BN}, but with a angle dependent $A_1(\theta)$, therefore
\begin{equation}
V(x,y,d,\theta) = k_e\frac{2A_1(\theta)q}{a}e^{\frac{-4\pi d}{\sqrt{3}a}}
\end{equation} 
where
\begin{equation}
	A_1(\theta)=-2\sqrt{3}\sin(2\pi y(\theta)-\frac{\pi}{3})\left[\cos(2\pi x(\theta))+\cos(2\pi y(\theta)-\frac{\pi}{3})\right] \nonumber
\end{equation}
and $x(\theta)= x\cos \theta +\frac{b}{a}y\sin \theta, y(\theta)=-\frac{a}{b}x\sin \theta + y\cos \theta$
%\begin{align}
%	A_1(\theta) = -2\sqrt{3}\sin\left[2\pi (-x\frac{a}{b}\sin \theta + y\cos \theta)-\frac{\pi}{3}\right] \left\{\cos\left[2\pi(x\cos \theta +\frac{by}{a}\sin \theta) \right]  \right. \nonumber  \\
%	 \left. + \cos\left[2\pi(-\frac{ax}{b}\sin \theta  y \cos \theta)-\frac{\pi}{3}\right] \right\}
%\end{equation}

\subsection{Formula Derivation for Hexagonal BN, AlN and GaN with Buckling}

For monolayer hexagonal structures, we consider here the case when the atoms are not planar, i.e.two point charges $q$ at position $\bs r(x,y,z)$ of  $(-1/2a,-1/b,z_1), (0,0,z_2)$ and two N atoms with charge $-q$ at $(-1/2a, -1/6b,z_3), (0,1/3b, z_4)$. when defining the average position of all the atoms in $z$ direction as $z=0$, we have:
\begin{align}\label{eq:BNz}
V(x,y,-d) 
%& =  \frac{q}{4\pi ab} \sum_{\vec{k}}{}' \left\{\frac{e^{-2\pi (d-z_1) M_k}}{M_k} \cos[2\pi (k_1x+k_2y)] + \frac{e^{-2\pi (d-z_2)M_k}}{M_k} \cos[2\pi(k_1(x-\frac{1}{2})+k_2(y-\frac{1}{2})] \right\}\nonumber \\
%& -\frac{q}{4\pi ab}\sum_{\vec{k}}{}'  \frac{e^{-2\pi (d-z_3) M_k}}{M_k} \cos[2\pi(k_1x+k_2(y-\frac{1}{3}))] \\
%&  + \frac{q}{4\pi ab}\sum_{\vec{k}}{}'  \frac{e^{-2\pi (d-d_4)M_k}}{M_k} \cos[2\pi k_1(x-\frac{1}{2})+2\pi k_2(y-\frac{5}{6})] \nonumber \\
& = k_e \frac{q}{ab}\sum_{\vec{k}}{}' cos[2\pi(k_1x+k_2y)] \left((-1)^{k_1+k_2} e^{-2\pi z_1 M_k} +  e^{-2\pi z_2M_k}\right) \frac{e^{-2\pi dM_k}}{M_k} \nonumber \\
& - k_e \frac{q}{ ab}\sum_{\vec{k}}{}' cos[2\pi k_1x + 2\pi k_2(y-\frac{1}{3})] \left( (-1)^{k_1+k_2} e^{-2\pi z_3 M_k} +e^{-2\pi z_4M_k}\right) \frac{e^{-2\pi dM_k}}{M_k} \nonumber \\ 
& + k_e\frac{q}{ab}(z_1+z_2-z_3-z_4)
\end{align}
where the last term in Eq.~\ref{eq:BNz} is correcting for the dipole effects. We then can derive the analytical formula for two different cases here: 

\noindent \textbf{Case \Romannum{1}: $z_1=z_2=-z_3=-z_4=\Delta z$,}
\begin{align}
%full
%V_1(x,y,z=d) = & \frac{4q}{4\pi} \frac{e^{\frac{-4\pi d}{\sqrt{3}a}} }{2a^2} \{ [\cos(2\pi x+2\pi y)+\cos(2\pi x-2\pi y)+cos(4\pi y)] e^{-4\pi \Delta z/(\sqrt{3}a) }  \nonumber \\
%& - \left[\cos(2\pi x+2\pi y -\frac{2\pi}{3}) +\cos(2\pi x-2\pi y + \frac{2\pi}{3}) + \cos(4\pi y -\frac{4\pi}{3}) \right] e^{4\pi \Delta z/(\sqrt{3}a)} \}\\
%
%V(x,y,z=d)=	&\frac{4\sqrt{3}q}{4 \pi a} \sin(2\pi y-\frac{\pi}{3})\left[ \cos(2\pi x) + \cos(2\pi y-\frac{\pi}{3})\right] e^{\frac{-4\pi d}{\sqrt{3}a}} \nonumber \\
%	& - \frac{4\pi \Delta zq}{4\pi a} \left[2\cos(2\pi x)\cos(2\pi y-\frac{\pi}{3}) - \cos(4\pi y - \frac{2\pi}{3})\right]e^{\frac{-4\pi d}{\sqrt{3}a}} + \frac{4q\Delta z}{4 \pi \sqrt{3}a^2}
V(x,y,z=-d) &=  k_e \frac{q}{ab}\sum_{\vec{k}}{}' \frac{e^{-2\pi d M_k}}{M_k}
\{ 
\cos\left[2\pi (k_1x+k_2y)\right]e^{-2\pi \Delta zM_k} -  
\cos\left[2\pi (k_1x+k_2(y-\frac{1}{3}))\right] e^{2\pi \Delta zM_k} 
\} %\nonumber \\
%& \approxeq k_e \frac{2q}{a}e^{-\frac{4\pi d}{\sqrt{3}a}} 
%\{
%  A_1 e^{-\frac{4\pi \Delta z}{\sqrt{3}a}} 
%  - A_2 e^{\frac{4\pi \Delta z}{\sqrt{3}a}}
%\} 
%& k_e\frac{q}{a} e^{\frac{-4\pi d}{\sqrt{3}a}} \{\left[4\cos(2\pi x) \cos(2\pi y)+\cos(4\pi y) \right] e^{\frac{-4\pi \Delta z}{\sqrt{3}a}} \nonumber \\
%& +\left[4\cos(2\pi x)\cos(2\pi y - \frac{2\pi}{3})+\cos(4\pi y-\frac{4\pi}{3}) \right] e^{\frac{4\pi \Delta z}{\sqrt{3}a}} \} + k_e \frac{4q\Delta z}{\sqrt{3}a^2}
\end{align}
%where $A_1=\left[2\cos(2\pi x) \cos(2\pi y) + \cos(4\pi y)\right], A_2=\left[2\cos(2\pi x) \cos(2\pi y - \frac{2\pi}{3}) + \cos(4\pi y - \frac{4\pi}{3}) \right]$. 
Using Taylor expansion, we can achieve $>95$\% accuracy to the first order, and $>$99\% accuracy to the second order (Eq.~\ref{eq:z195}), therefore
%\begin{align}
%V(x,y,z=-d)=& k_e \frac{2q}{a}e^{-\frac{4\pi d}{\sqrt{3}a}} 
%\{
%A_1 e^{-\frac{4\pi \Delta z}{\sqrt{3}a}} 
%- A_2 e^{\frac{4\pi \Delta z}{\sqrt{3}a}}
%\} \\
%\approxeq & 
%\begin{cases}
%k_e\frac{2q}{a}\left[ (A_1-A_2)(1+\frac{4\pi^2 \Delta z^2}{3a^2}) - (A_1+A_2)\frac{4\pi \Delta z}{\sqrt{3}a} \right] e^{-\frac{4\pi d}{\sqrt{3}a}} \\%  \label{eq:z199} \\
%k_e\frac{2q}{a}\left[ (A_1-A_2)- (A_1+A_2)\frac{4\pi \Delta z}{\sqrt{3}a} \right] e^{-\frac{4\pi d}{\sqrt{3}a}}  \label{eq:z195} 
% \end{cases} 
%\end{align}
\begin{align}
V(x,y,z=-d)=& k_e \frac{q}{a}e^{-\frac{4\pi d}{\sqrt{3}a}} 
\{
(A_1+A_1^{\prime}) e^{-\frac{4\pi \Delta z}{\sqrt{3}a}} 
- (A_1^{\prime}-A_1) e^{\frac{4\pi \Delta z}{\sqrt{3}a}}
\} \\
\approxeq & 
\begin{cases}
k_e\frac{2q}{a}\left[ A_1(1+\frac{8\pi^2 \Delta z^2}{3a^2}) - A_1^{\prime}\frac{4\pi \Delta z}{\sqrt{3}a} \right] e^{-\frac{4\pi d}{\sqrt{3}a}} \\%  \label{eq:z199} \\
k_e\frac{2q}{a}\left[ A_1- A_1^{\prime}\frac{4\pi \Delta z}{\sqrt{3}a} \right] e^{-\frac{4\pi d}{\sqrt{3}a}}  \label{eq:z195} 
\end{cases} 
%& V(x,y,z=-d)=k_e\frac{2q}{a}\left[ (A_1-A_2)+ (A_1+A_2)\frac{4\pi \Delta z}{\sqrt{3}a} \right] e^{-\frac{4\pi d}{\sqrt{3}a}}  \label{eq:z195} \\
%& V(x,y,z=-d)=k_e\frac{2q}{a}\left[ (A_1-A_2)(1+\frac{4\pi^2 \Delta z^2}{3a^2}) + (A_1+A_2)\frac{4\pi \Delta z}{\sqrt{3}a} \right] e^{-\frac{4\pi d}{\sqrt{3}a}}  \label{eq:z199}
\end{align}
%where $A_1+A_2=2\left[2\cos(2\pi x) \cos(2\pi y) + \cos(4\pi y)\right], A_2-A_1=2\left[2\cos(2\pi x) \cos(2\pi y - \frac{2\pi}{3}) + \cos(4\pi y - \frac{4\pi}{3}) \right]$.
where $A_1 = -2\sqrt{3}\sin(2\pi y-\frac{\pi}{3})\left[\cos(2\pi x)+\cos(2\pi y-\frac{\pi}{3}) \right]$, and\\
$A_1^{\prime} = 2\cos(2\pi y-\frac{\pi}{3})\left[\cos(2\pi x)-\cos(2\pi y-\frac{\pi}{3}) \right]+1$. 

%\begin{empheq}[left={
%	V(x,y,z=-d)\approxeq \empheqlbrace}]{alignat=1}
%& k_e\frac{2q}{a}\left[ (A_1-A_2)(1+\frac{4\pi^2 \Delta z^2}{3a^2}) + (A_1+A_2)\frac{4\pi \Delta z}{\sqrt{3}a} \right] e^{-\frac{4\pi d}{\sqrt{3}a}}   \label{eq:z199} \\
%& k_e\frac{2q}{a}\left[ (A_1-A_2)+ (A_1+A_2)\frac{4\pi \Delta z}{\sqrt{3}a} \right] e^{-\frac{4\pi d}{\sqrt{3}a}}  \label{eq:z195} 
%\end{empheq}

\vspace{12pt}
\noindent \textbf{Case \Romannum{2}: $z_1=-z_2=-z_3=z_4=\Delta z$,}
\begin{align}
	V(x,y,z=-d) = &k_e\frac{q}{ab} \sum_{\vec{k}}{}' \frac{e^{-2\pi dM_k}}{M_k} \left[e^{2\pi \Delta zM_k} + (-1)^{k_1+k_2}e^{-2\pi \Delta zM_k} \right] \nonumber \\
	& \left\{ \cos \left[2\pi (k_1x+ k_2 y)\right]-(-1)^{k_1+k_2} \cos \left[2\pi(k_1x+k_2(y-\frac{1}{3})) \right] \right\}
\end{align}
We can achieve $>99$\% accuracy with $-1\leq k_1 \leq 1, -2 \leq k_2 \leq 2$ and to second order Taylor expansion, therefore
%\begin{align}
%	V(x,y,z=-d) = k_e \frac{q}{\sqrt{3} a^2}\left[
%	\sqrt{3}a B_1 \left(1+\frac{8\pi^2 \Delta z^2}{3a^2}\right)e^{\frac{-4\pi d}{\sqrt{3}a}} + 4\pi \Delta z B_2 e^{\frac{-2\pi d}{\sqrt{3}a}}  \right.  \nonumber \\
%	\left. + 4\pi \Delta zB_3e^{\frac{-2\pi d}{a}}  +4\pi \Delta zB_4 e^{\frac{-2\sqrt{7}\pi d}{\sqrt{3}a}} \right]
%\end{align}
\begin{align}
V(x,y,z=-d) = k_e \frac{2q}{a}\left[A_1 \left(1+\frac{8\pi^2 \Delta z^2}{3a^2}\right)e^{\frac{-4\pi d}{\sqrt{3}a}} + A_2 \frac{\Delta z}{a} e^{\frac{-2\pi d}{\sqrt{3}a}}  + A_3 \frac{\Delta z}{a}e^{\frac{-2\pi d}{a}}  +A_4 \frac{\Delta z}{a} e^{\frac{-2\sqrt{7}\pi d}{\sqrt{3}a}} \right]
\end{align}

and $>98$\% accuracy to the first order in Taylor expansion, where
%\begin{align}
%	V(x,y,z=-d) = k_e \frac{q}{\sqrt{3} a^2}\left[
%\sqrt{3}a B_1 e^{\frac{-4\pi d}{\sqrt{3}a}} +  4\pi \Delta z B_2 e^{\frac{-2\pi d}{\sqrt{3}a}} + 4\pi \Delta zB_3 e^{\frac{-2\pi d}{a}} + 4\pi \Delta zB_4 e^{\frac{-2\sqrt{7}\pi d}{\sqrt{3}a}}  \right]
%\end{align}
\begin{align}
V(x,y,z=-d) = k_e \frac{2q}{a}\left[A_1e^{\frac{-4\pi d}{\sqrt{3}a}} + A_2 \frac{\Delta z}{a} e^{\frac{-2\pi d}{\sqrt{3}a}}  + A_3 \frac{\Delta z}{a}e^{\frac{-2\pi d}{a}}  +A_4 \frac{\Delta z}{a} e^{\frac{-2\sqrt{7}\pi d}{\sqrt{3}a}} \right]
\end{align}

where $A_2=4\pi \cos(2\pi y-\frac{\pi}{3})/\sqrt{3}, A_3 = 8\pi \cos(2\pi x)/\sqrt{3}, A_4=8\pi \cos(2\pi x)\cos(4\pi y-\frac{2\pi}{3})$.

%\begin{align}
%V(x,y,z=d)=	&\frac{4\sqrt{3}q}{4 \pi a} \sin(2\pi y-\frac{\pi}{3})\left[ \cos(2\pi x) + \cos(2\pi y-\frac{\pi}{3})\right] e^{\frac{-4\pi d}{\sqrt{3}a}} \nonumber \\
%& - \frac{4\pi \Delta zq}{4\pi a} \left[4\cos(2\pi x)\sin^2(2\pi y-\frac{\pi}{3}) + \cos(2\pi y - \frac{\pi}{3})\right]e^{\frac{-4\pi d}{\sqrt{3}a}} 
%\end{align}

\subsection{Formula For Metal MonoChalcogenides}

\begin{figure}[H]
	\centering
	\includegraphics[width=0.4\linewidth]{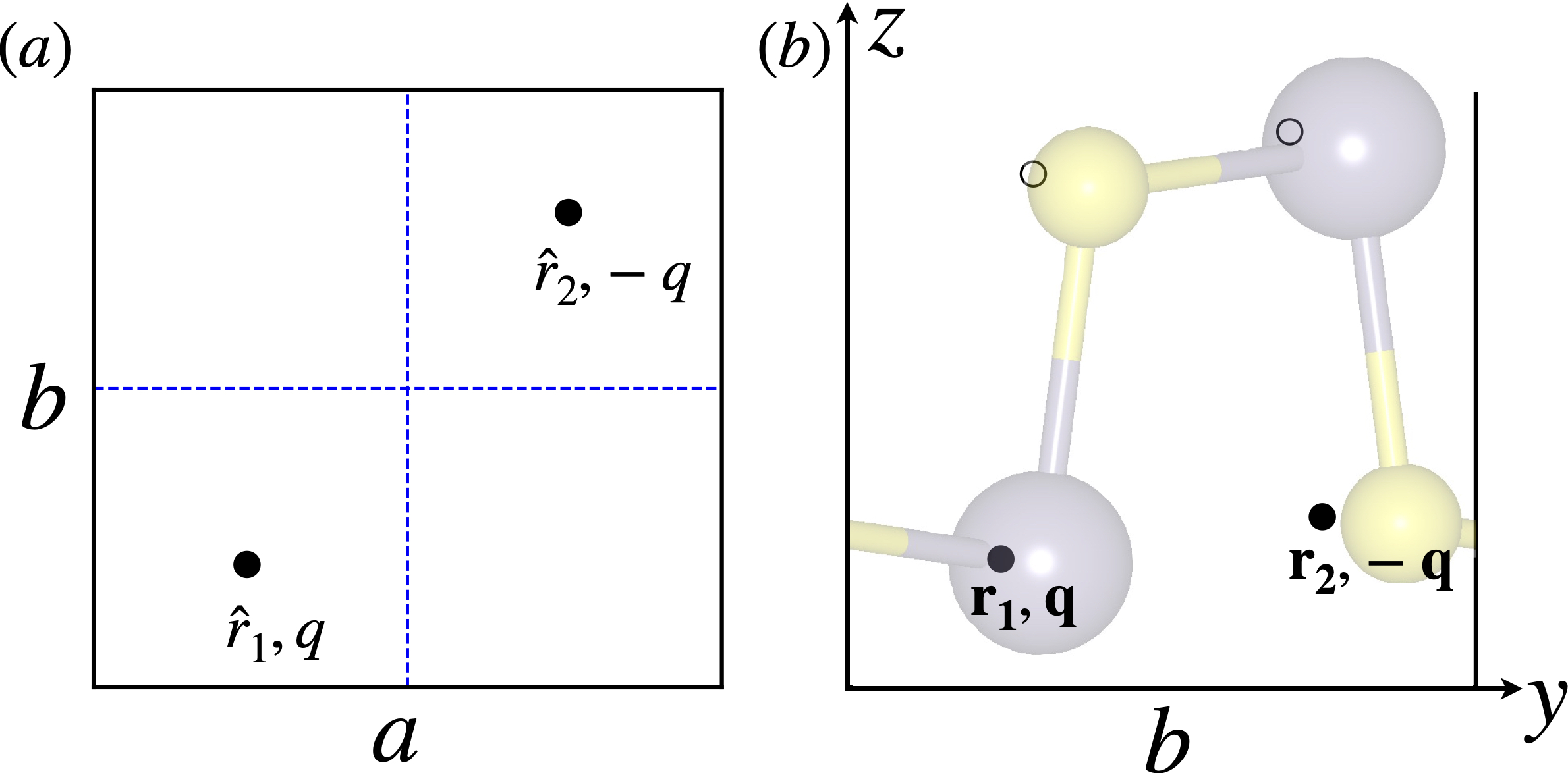}
	\caption{(a) top view and (b) side view of the unit cell and the positions of the point charges for metal monochalcogenides shown as black dots.} \label{fig:MMCq}
\end{figure}

Monolayer of metal monochalcogenides (MMC) MX where M=Ge, Sn, Pb and X=S,Se,Te has four atoms per unit cell as shown in Fig.~\ref{fig:MMCq} (b). We can use two point charges $\hat{r}_1 (-x_0a, -y_0b, -\Delta z)$ and $\hat{r}_2 (x_0a, y_0b, \Delta z)$ with charges of $q$ and $-q$, respectively, to represent each MX as shown in Fig.~\ref{fig:MMCq}. %The minimum and maximum potentials are located at $\hat{r}_1$ and $\hat{r}_2$, respectively. 
Since the total dipole moment is zero, the potential is\\
\begin{align}
%\begin{split}
	V(x,y,z=d) &=k_e\sum_i \frac{q_i}{ab}  \sum_{\vec{k}}{}' \frac{e^{-2\pi (d-z_i)M_k }}{M_k} cos\left\{2\pi \left[k_1(x-x_i)+k_2(y-y_i)\right]\right\} \\ 
	& = k_e\frac{q}{ab} \sum_{\vec{k}}{}' \frac{e^{-2\pi dM_k}}{M_k} \left\{  \cos\{2\pi\left[k_1(x+x_0)+k_2(y+y_0)\right]\}e^{2\pi \Delta z M_k} \right.  \nonumber \\
	& \qquad \qquad \qquad \qquad  \left. {} - \cos\{2\pi\left[k_1(x-x_0)+k_2(y-y_0)\right]\}e^{-2\pi \Delta z M_k}
	\right\}	
%\end{split}	
%	
%	  \frac{e^{2\pi \Delta z M_k}e^{-2\pi dM_k}}{M_k}  \nonumber \\
%	&  - k_e\frac{q}{ab} \sum_{\vec{k}}{}' \cos\left\{2\pi\left[k_1(x-x_0)+k_2(y-y_0)\right]\right\} \frac{e^{-2\pi \Delta z M_k}e^{-2\pi dM_k}}{M_k}
\end{align}

As we can achieve accuracy $> 98$\% with $-1 \leq k_1, k_2  \leq 1$, therefore
%\begin{equation}
\begin{align}\label{eq:mmc}
%\begin{split}
V(x,y,z=d) = & k_e \frac{q}{a} \left\{ \frac{4\left[C_x C_y e^{2\pi \Delta z \sqrt{1/a^2+1/b^2}} - D_x D_y e^{-2\pi \Delta z\sqrt{1/a^2+1/b^2}} \right]}{\sqrt{b^2/a^2+1}} e^{-2\pi d\sqrt{1/a^2+1/b^2}} \nonumber \right. \\ 
& \left. {\qquad} + 2  \left[C_y e^{2\pi \Delta z/b} - D_y e^{-2\pi \Delta z/b} \right]e^{\frac{-2\pi d}{b}} \right. \nonumber \\
& \left. {\qquad} + \frac{2a }{b} \left[C_x e^{2\pi \Delta z/a} - D_x e^{-2\pi \Delta z/a} \right]e^{\frac{-2\pi d}{a}}  \right\}
%& \left.{\qquad} - \cos(2\pi \Delta x_2)\cos(2\pi \Delta y_2)e^{-2\pi \Delta z \sqrt{1/a^2+1/b^2}} \right] \nonumber \\ 
% & \left. {\qquad} + \frac{2e^{-2\pi d/a}}{b} \left[ \cos(2\pi \Delta x_1) e^{2\pi \Delta z/a} - \cos(2\pi \Delta x_2) e^{-2\pi \Delta z/a}\right] \nonumber  \right. \\
% & \left. {\qquad} + \frac{2e^{-2\pi d/b}}{a} \left[\cos(2\pi \Delta y_1) e^{2\pi \Delta z/b} - \cos(2\pi \Delta y_2)e^{-2\pi \Delta z/b} \right]
% \right\}
%
%\frac{2q}{4\pi ab} \nonumber & \{ \frac{e^{-2\pi d \sqrt{1/a^2+1/b^2}}}{\sqrt{1/a^2+1/b^2}} 2\left[ \cos(2\pi \Delta x_1)\cos(2\pi \Delta y_1) - \cos(2\pi \Delta x_2)\cos(2\pi \Delta y_2)\right] \nonumber \\
% & + ae^{-2\pi d/a} \left[\cos(2\pi \Delta x_1) - \cos(2\pi \Delta x_2)\right] \nonumber \\
% & + be^{-2\pi d/b} \left[\cos(2\pi \Delta y_1)-\cos(2\pi \Delta y_2)\right]\} \\
\end{align}
%where $\Delta x_1 = x+x_0, \Delta x_2 = x-x_0, \Delta y_1 = y+y_0, \Delta y_2 = y-y_0$.  
where $C_x=\cos[2\pi (x+x_0)], D_x=\cos[2\pi (x-x_0)],C_y=\cos[2\pi (y+y_0)],D_y=\cos[2\pi (y-y_0)]$.
We can achieve $>98$\% accuracy using Taylor expansion to the second order (Eq.~\ref{eq:mmc2}), and $>95$\% to the first order(Eq.~\ref{eq:mmc1}), we have
\begin{align} \label{eq:mmc2}
%%	V(x,y,z=d) =& k_e q \left\{ \frac{4e^{-2\pi d \sqrt{1/a^2+1/b^2}}}{\sqrt{a^2+b^2}} \left[\cos(2\pi \Delta x_1) \cos(2\pi \Delta y_1) - \cos(2\pi \Delta x_2) \cos(2\pi \Delta y_2) \right] \nonumber  \right. \\
%% & \left. {\qquad} + \frac{2e^{-2\pi d/a}}{b} \left[ \cos(2\pi \Delta x_1)  + \cos(2\pi \Delta x_2) \right] \nonumber  \right. \\
%%& \left. {\qquad} + \frac{2e^{-2\pi d/b}}{a} \left[\cos(2\pi \Delta y_1) + \cos(2\pi \Delta y_2)\right]
%%\right\}
%V(x,y,z=d) = & k_e \frac{q}{a} \left\{ \frac{B_1\left[1+4\pi^2 (\frac{b^2}{a^2}+1)(\frac{\Delta z}{b})^2\right] + B_1^{\prime} 2\pi \sqrt{\frac{b^2}{a^2}+1}\frac{\Delta z}{b} }{\sqrt{b^2/a^2+1}} e^{-2\pi d\sqrt{1/a^2+1/b^2}} \nonumber \right. \\ 
%& \left. {\qquad} + \left[B_2(1+(2\pi \frac{\Delta z}{b})^2)+B_2^{\prime}2\pi \frac{\Delta z}{b}\right]e^{\frac{-2\pi d}{b}} \right. \nonumber \\
%& \left. {\qquad} +  \frac{a}{b}\left[B_3 (1+ (2\pi \frac{\Delta z}{a})^2)+B_3^{\prime} 2\pi\frac{\Delta z}{a}\right] e^{\frac{-2\pi d}{a}} \right\} 
V(x,y,d) & = k_e \frac{q}{a}  \left\{ B_1(x,y,\Delta z) e^{-2\pi d\sqrt{1/a^2+1/b^2}} \nonumber \right. \\
& \left. + B_2(x,y,\Delta z) e^{\frac{-2\pi d}{b}} + B_3(x,y,\Delta z) e^{\frac{-2\pi d}{a}}  \right\}
\end{align}
\begin{align} \label{eq:mmc1}
V(x,y,z=d) = & k_e \frac{q}{a} \left\{ \frac{B_1+ B_1^{\prime} 2\pi \sqrt{\frac{b^2}{a^2}+1}\frac{\Delta z}{b} }{\sqrt{b^2/a^2+1}} e^{-2\pi d\sqrt{1/a^2+1/b^2}} \nonumber \right. \\ 
& \left. {\quad} + \left[B_2+B_2^{\prime}2\pi \frac{\Delta z}{b}\right]e^{\frac{-2\pi d}{b}} +  \frac{a}{b}\left[B_3+B_3^{\prime} 2\pi\frac{\Delta z}{a}\right] e^{\frac{-2\pi d}{a}} \right\} 
\end{align}
where $B_1 = 4(C_xC_y-D_xD_y), B_1^{\prime}=4(C_xC_y+D_xD_y), B_2=2(C_y-D_y), B_2^{\prime}=2(C_y+D_y), B_3 = 2(C_x-D_x), B_3^{\prime}= 2(C_x+D_x)$.

%Adding the second term in the Taylor expansion for $e^{\pm 2\pi \Delta z M_k}$, the accuracy can be improved by 0.1$\sim$3\% for all MMCs depending on the value of $\Delta z$. The corresponding potential adding to Eq.~\ref{eq:mmc} is:
%\begin{align}
%V(x,y,z=d) = \frac{4q \pi \Delta z}{4\pi ab} \nonumber & \{ 2e^{-2\pi d \sqrt{1/a^2+1/b^2}} \left[ \cos(2\pi \Delta x_1)\cos(2\pi \Delta y_1) + \cos(2\pi \Delta x_2)\cos(2\pi \Delta y_2)\right] \nonumber \\
%& + e^{-2\pi d/a} \left[\cos(2\pi \Delta x_1) + \cos(2\pi \Delta x_2)\right] \nonumber \\
%& + e^{-2\pi d/b} \left[\cos(2\pi \Delta y_1)+\cos(2\pi \Delta y_2)\right]\} \\
%\end{align}
%
%The minimum and maximum electrostatic potentials are located below the charges ($\hat{r_1}, q$) and ($\hat{r_2}, -q$), thereforre the in-plane potential modulation is,
%\begin{align}
%	\Delta V &= V_{max}(x_0a,y_0b,d) -V_{min}(-x_0a,-y_0b,d) \nonumber \\
%	&=  \frac{4q}{4\pi ab} \left\{ \frac{2[cos(4\pi x_0) cos(4\pi y_0)-1]}{\sqrt{1/a^2+1/b^2}} e^{-2\pi d \sqrt{\frac{1}{a^2}+\frac{1}{b^2}}}  % \nonumber \\
%	+  a[cos(4\pi x_0)-1]e^{-2\pi d/a} + b[cos(4\pi y_0)-1]e^{-2\pi d/b}\right\}
%\end{align}

\subsection{Analytical Formula for Moir\'{e} Structures}
For a unit cell rotated by angle $\theta$, the potential can be computed by replacing $x,y$ in Eq.\ref{eq:BN} and Eq.\ref{eq:mmc2} by (note that $x,y$ are in fractional coordinates of the cell vectors):
\begin{align}
    & x(\theta)=x\cos \theta + \frac{b}{a}y\sin \theta \nonumber \\
    & y(\theta) = -\frac{a}{b}x sin \theta + y\sin \theta
\end{align}
Therefore, by adding the potential with the above $x(\theta), y(\theta)$, we can obtain angle-specific analytical formula for moir\'{e} structures. For example, for two layers of h-BN in the configuration as shown in Fig. 4(a), the potential at the center 2D material can be computed as:
\begin{align}
    V(x,y,d_1,d_2,\theta)=k_e\frac{2q}{a}\left[A(x,y)e^{\frac{-4\pi d_1}{\sqrt{3}a}}+A(x(\theta),y(\theta))e^{\frac{-4\pi d_2}{\sqrt{3}a}}\right]
\end{align}
where $d_1, d_2$ are the distance of the two layers of h-BN, which are considered as equal in Fig.4. Same can be obtained for transition metal monochalcogenides using Eq.\ref{eq:mmc2}.

Meanwhile, using the analytical formula for the potential, we can derive their electric field by:
\begin{align}
    \vec{F}(x,y,z,\theta) = -\left[\frac{\partial V(x,y,d,\theta)}{\partial x} \vec{x} + \frac{\partial V(x,y,d,\theta)}{\partial y} \vec{y} +\frac{\partial V(x,y,d,\theta)}{\partial z} \vec{z}\right]
\end{align}

\section{Lattice Parameters and Charges}

\begin{table}[H]
	\centering
	\captionsetup{margin={1.0cm,1.0cm}}
	\caption{\ lattice parameters and the absolute magnitude of the point charges $q$ for each 2D material studied in this work. The positions for the positive and negative point charges are shown in Fig.3(d), Fig.\ref{figs:AlNz}(d), Fig.\ref{figs:GaNz}(d) for the hexagonal monolayer, and Fig.\ref{figs:MMCall} for MMCs.} 
	\label{tbl:qstr}
	\begin{tabular*}{0.7\textwidth}{@{\extracolsep{\fill}}cccc}
		\hline
		2D & a (\AA) & b (\AA) & $|q|$ \\ \hline %& $l_x/l_y$
		h-BN	& 2.51	& 2.51	& 0.89
 \\
		h-AlN	& 3.13	& 3.13	& 1.47
\\
		h-GaN	& 3.21	& 3.21	& 1.18
\\
	    GeS	& 3.65	& 4.50	& 0.23
\\
		GeSe	& 3.97	& 4.30	& 0.25
\\
		GeTe	& 4.39	& 4.24	& 0.18
\\
		SnS	& 4.09	& 4.30	& 0.26
\\
		SnSe	& 4.29	& 4.39	& 0.30
\\
		SnTe	& 4.55	& 4.58	& 0.25
\\

		PbS	& 4.25	& 4.25	& 0.50
\\
		PbSe	& 4.41	& 4.41	& 0.48
\\
		PbTe	& 4.64	& 4.64	& 0.36\\
		\hline
	\end{tabular*} 
\end{table}	

\section{Potential Results for h-AlN and h-GaN}
\begin{figure}[H]
	\centering
	\includegraphics[width=0.9\linewidth, center]{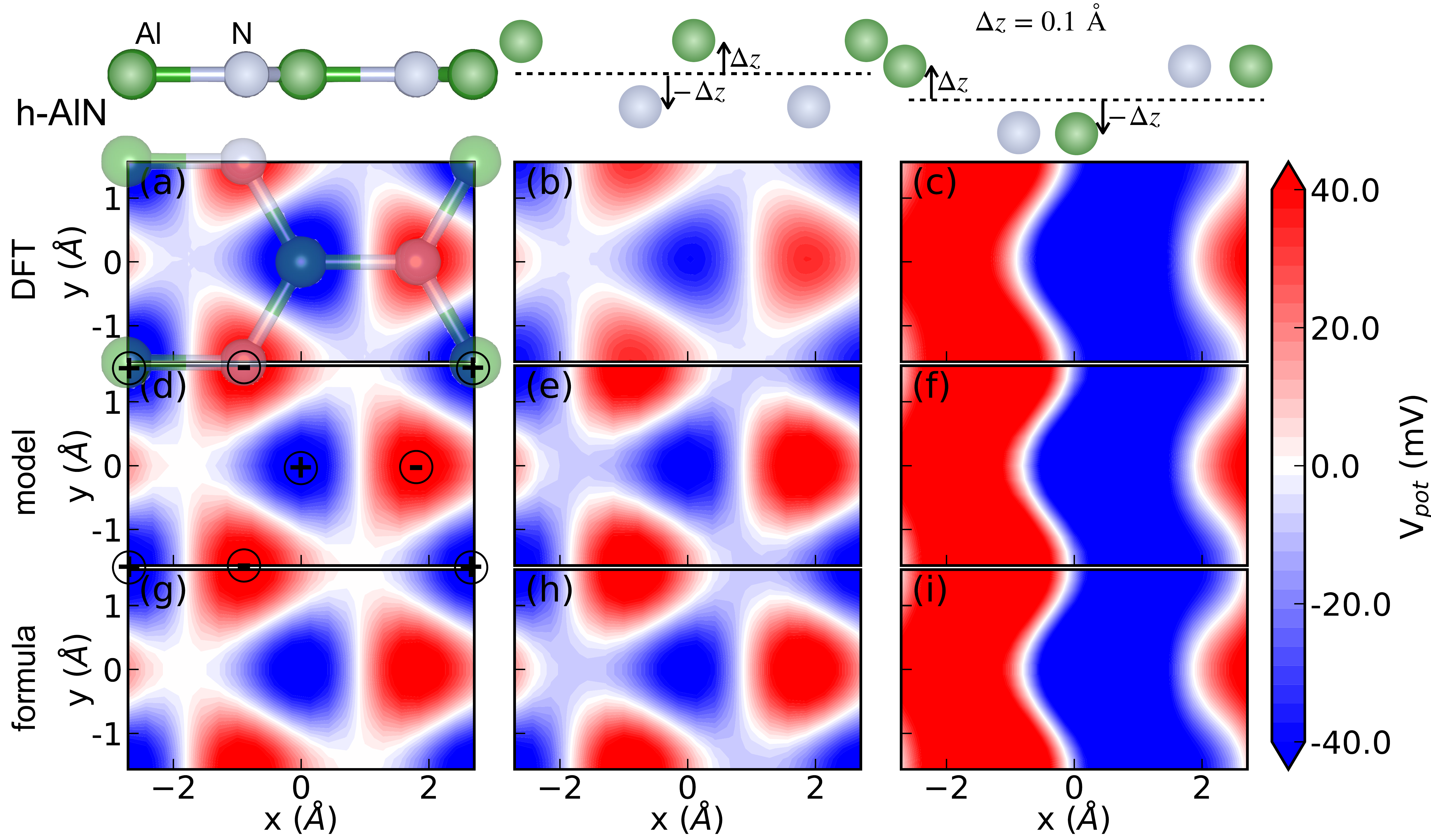}
	\caption{In-plane electrostatic potential at $d=3.03$ \AA from single layer of h-AlN computed from (a) DFT, (d) the DCD model, (g) formula. The left two columns show the potential for h-AlN with Raman-active (second column) and infrared-active (third column) phonon mode, see atomic configurations on the top, computed from DFT, DCD model and formula, respectively.} \label{figs:AlNz}
\end{figure}

\begin{figure}[H]
	\centering
	\includegraphics[width=0.9\linewidth, center]{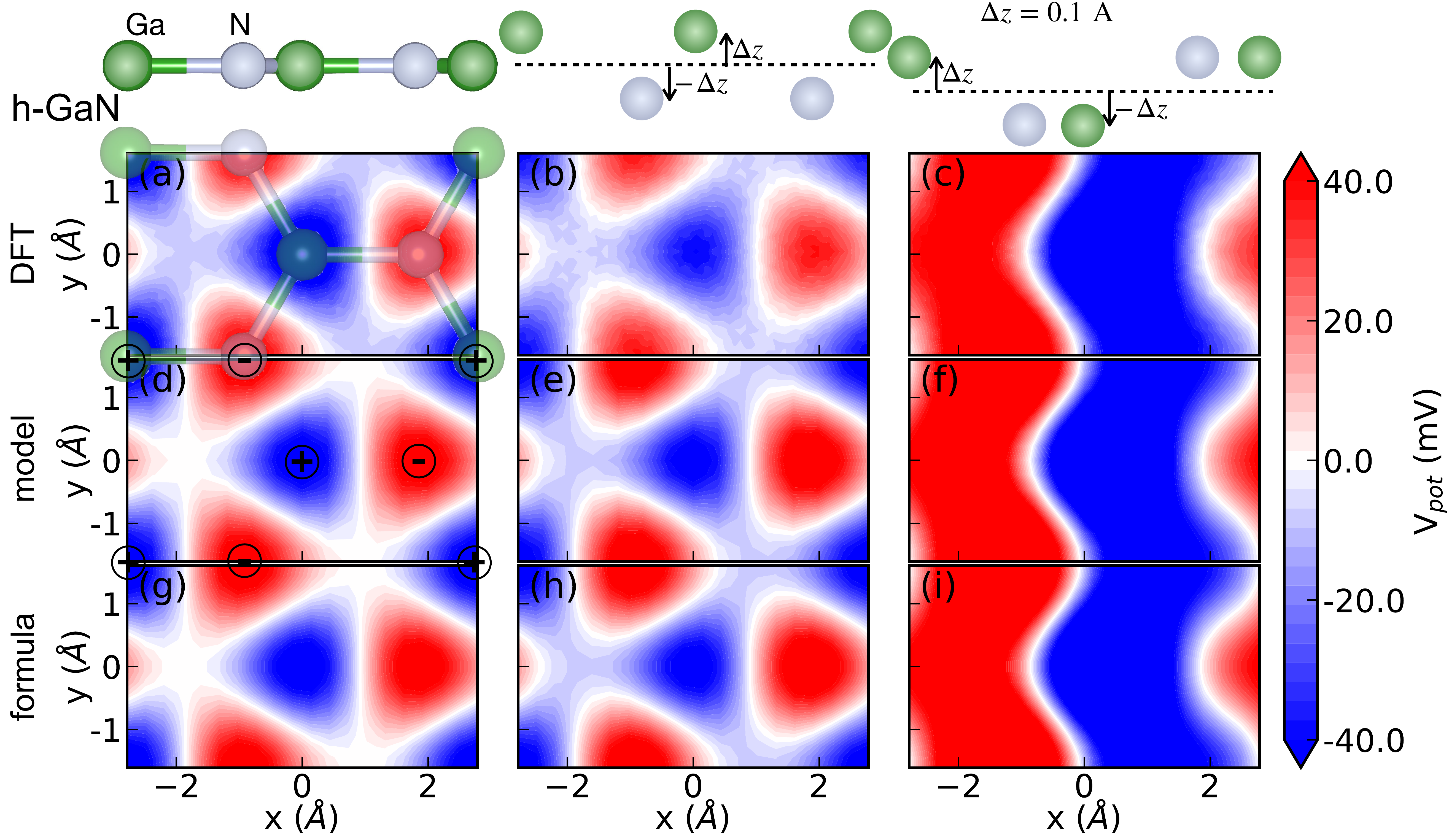}
	\caption{In-plane electrostatic potential at $d=3.03$ \AA from single layer of h-GaN computed from (a) DFT, (d) the DCD model, (g) formula. The left two columns show the potential for h-GaN with Raman-active (second column) and infrared-active (third column) phonon mode, see atomic configurations on the top, computed from DFT, DCD model and formula, respectively.} \label{figs:GaNz}
\end{figure}

\begin{figure}[H]
	\centering
	\includegraphics[width=0.9\linewidth, center]{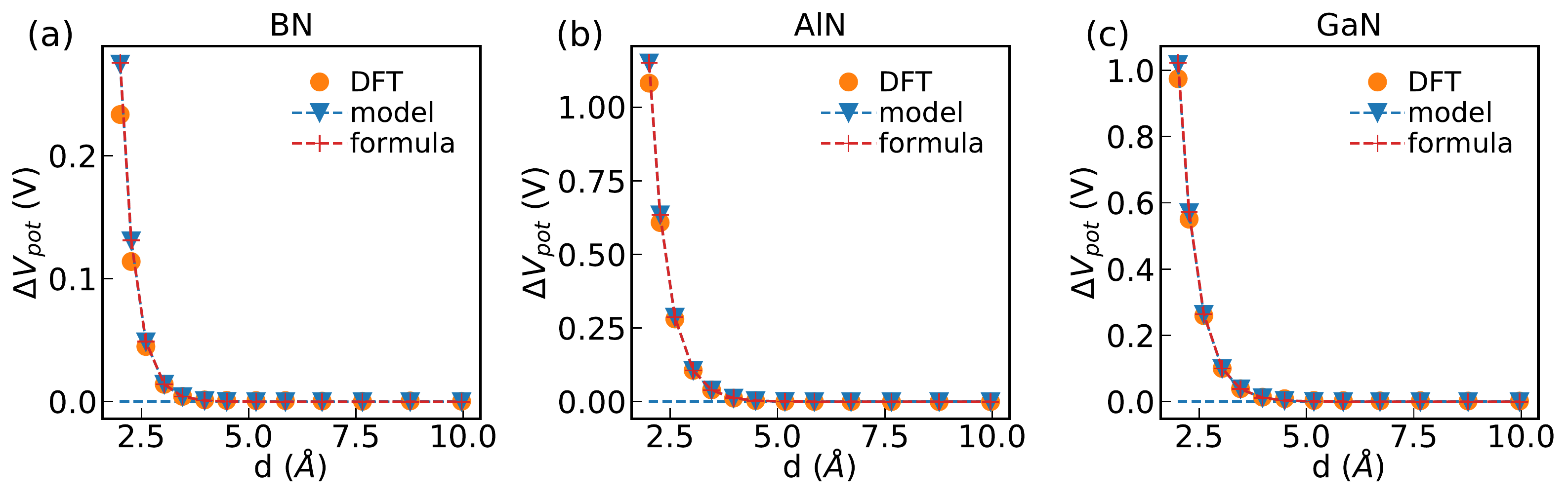}
	\caption{$\Delta V_{pot}=V_{max}-V_{min}$ as a function of $d$ for (a) h-BN, (b) h-AlN, (c) h-GaN computed from DFT, model and formula.} \label{figs:BNdline}
\end{figure}

\section{Potential Results for all Transition Metal MonoChalcogenides}
\begin{figure}[H]
	%\centering
	\subfloat{\includegraphics[width=0.99\linewidth,center]{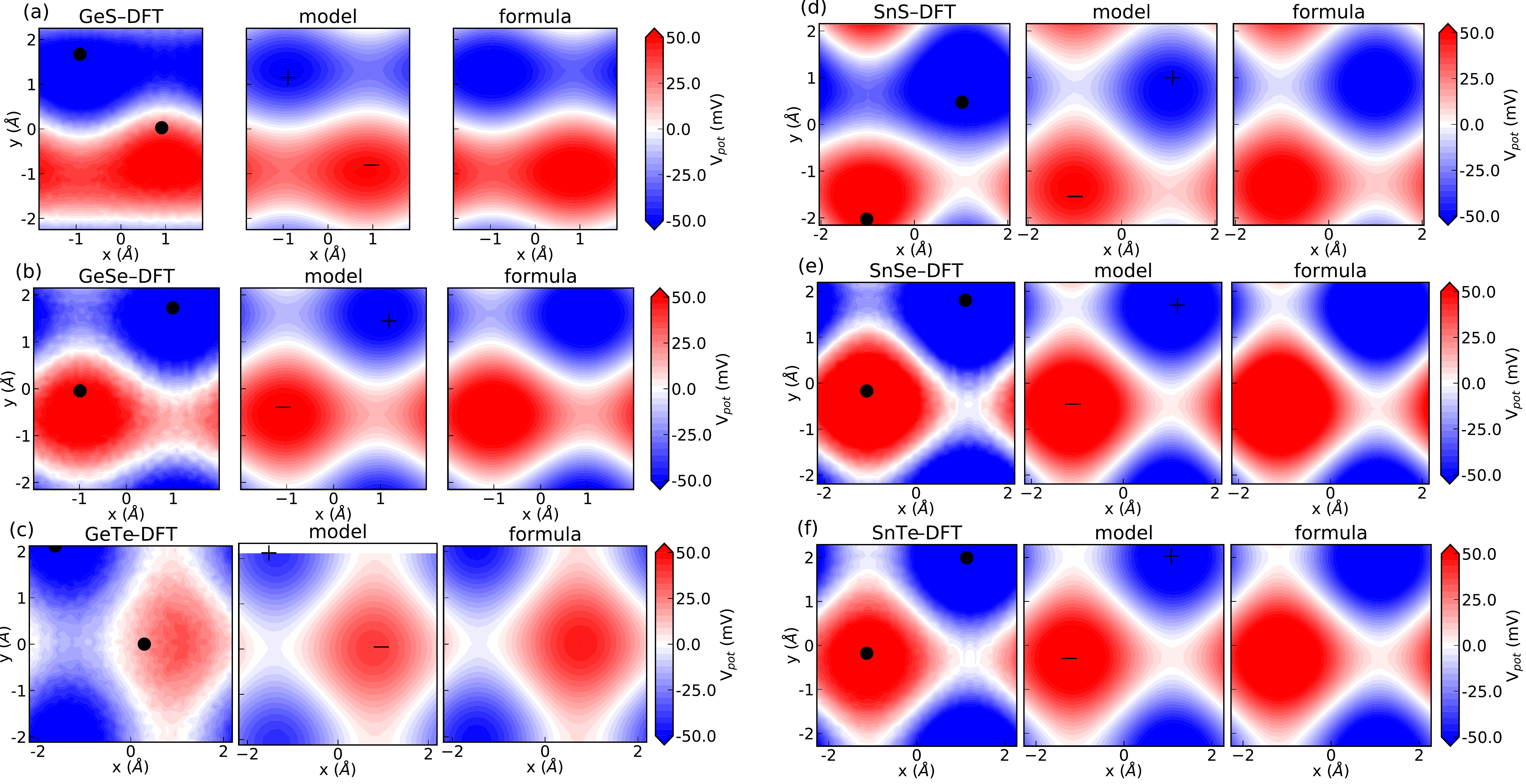}} \\%\hspace{0.5pt}
	\subfloat{\includegraphics[width=0.5\linewidth,center]{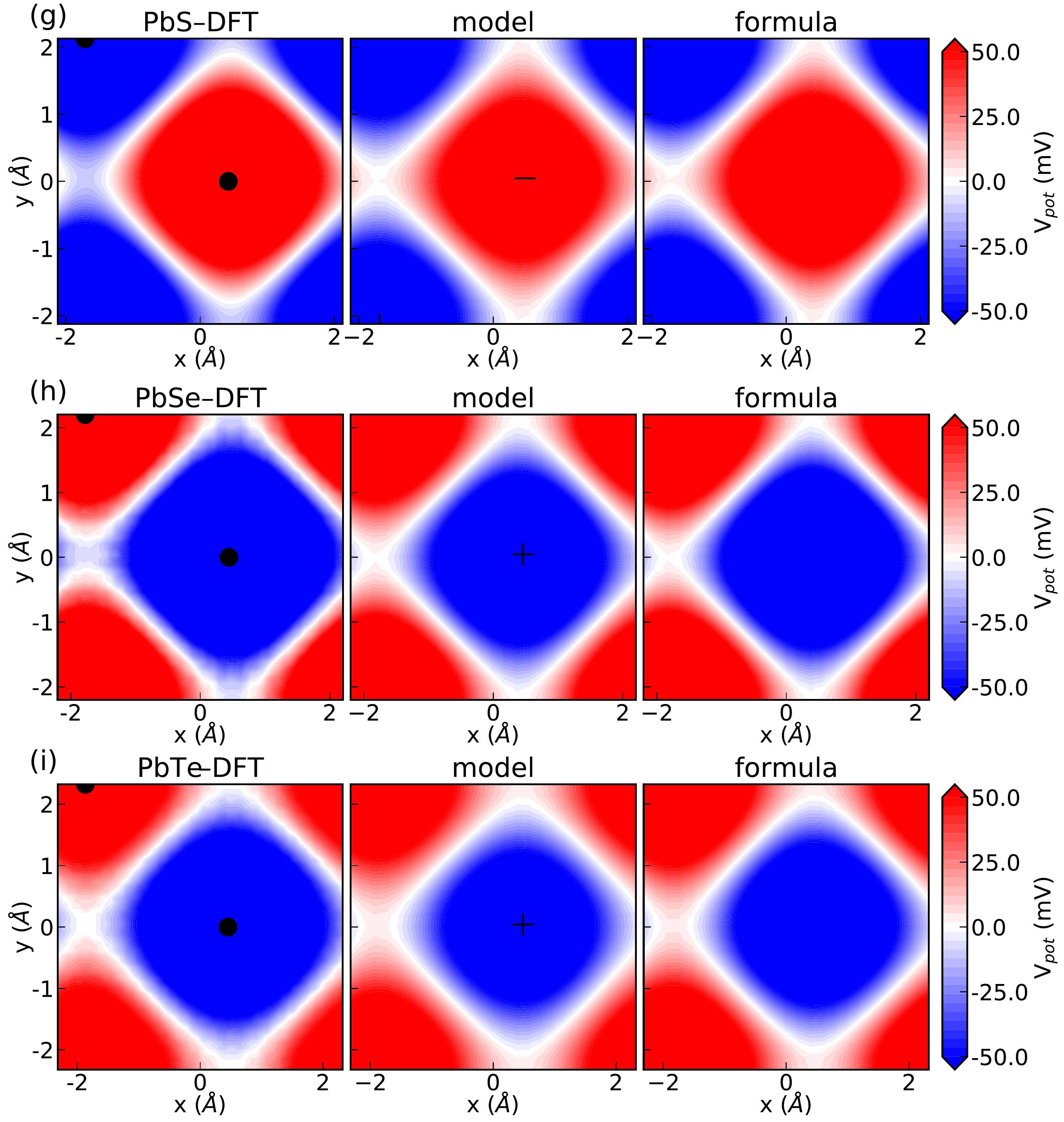}}
	\caption{Potential from transition metal monochalcogenides at distances of $\sim$ 3.1 \AA above computed using DFT, model and analytical formula in Eq.\ref{eq:mmc2}. The black dots show the positions of the bottom two atoms. The plus and minus symbols mark the positions of the postive and negative charges. } \label{figs:MMCall}
\end{figure}

\begin{figure}[H]
	\centering
	\includegraphics[width=0.9\linewidth, center]{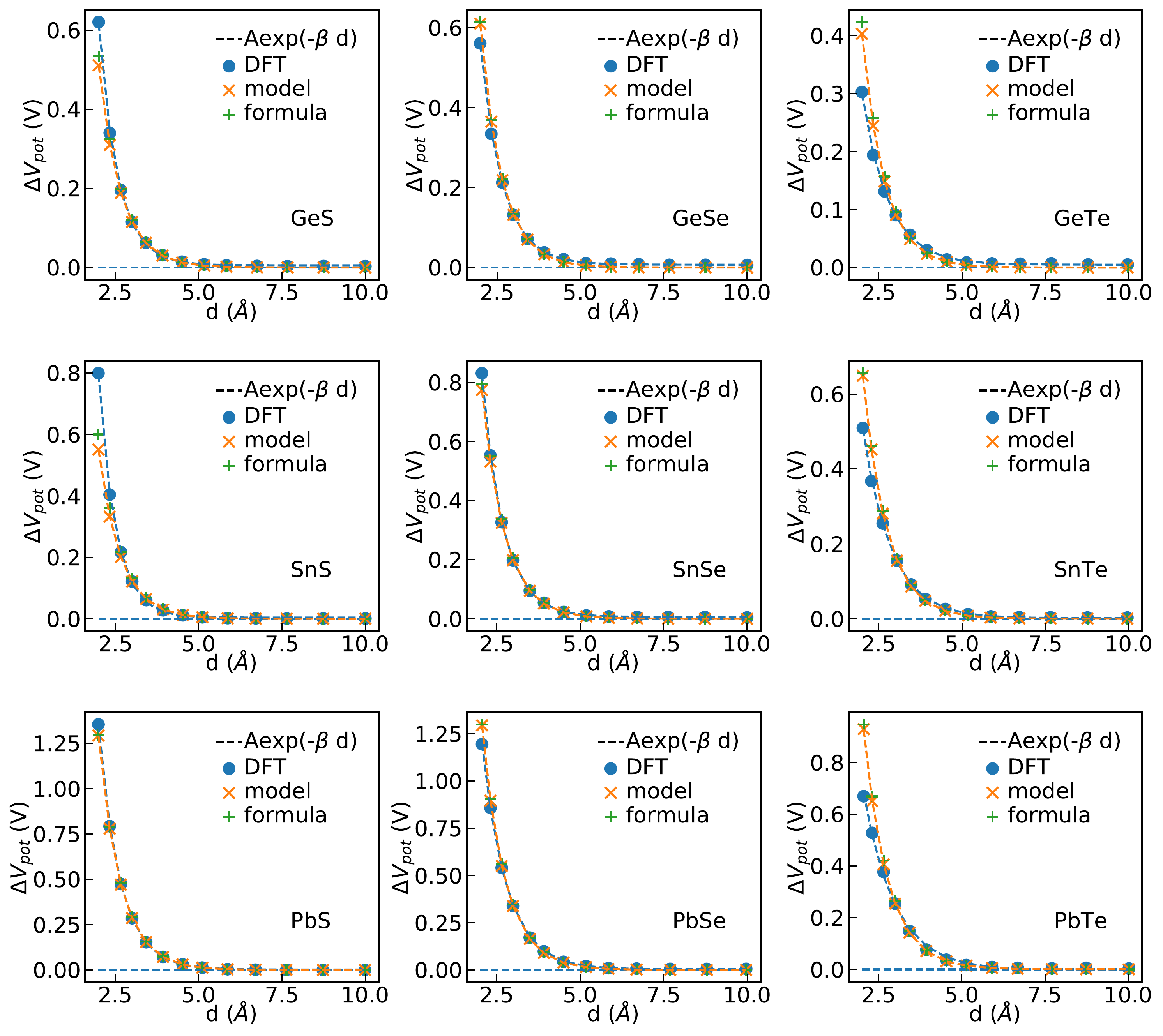}
	\caption{$\Delta V_{pot}=V_{max}-V_{min}$ as a function of $d$ for MMCs computed from DFT, model and formula.} \label{figs:MMCdline}
\end{figure}

\section{Impacts of Stacking and Twist Angle on the Potential and Electric Field for MMCs}
\begin{figure}[H]
	\centering
	\includegraphics[width=0.9\linewidth, center]{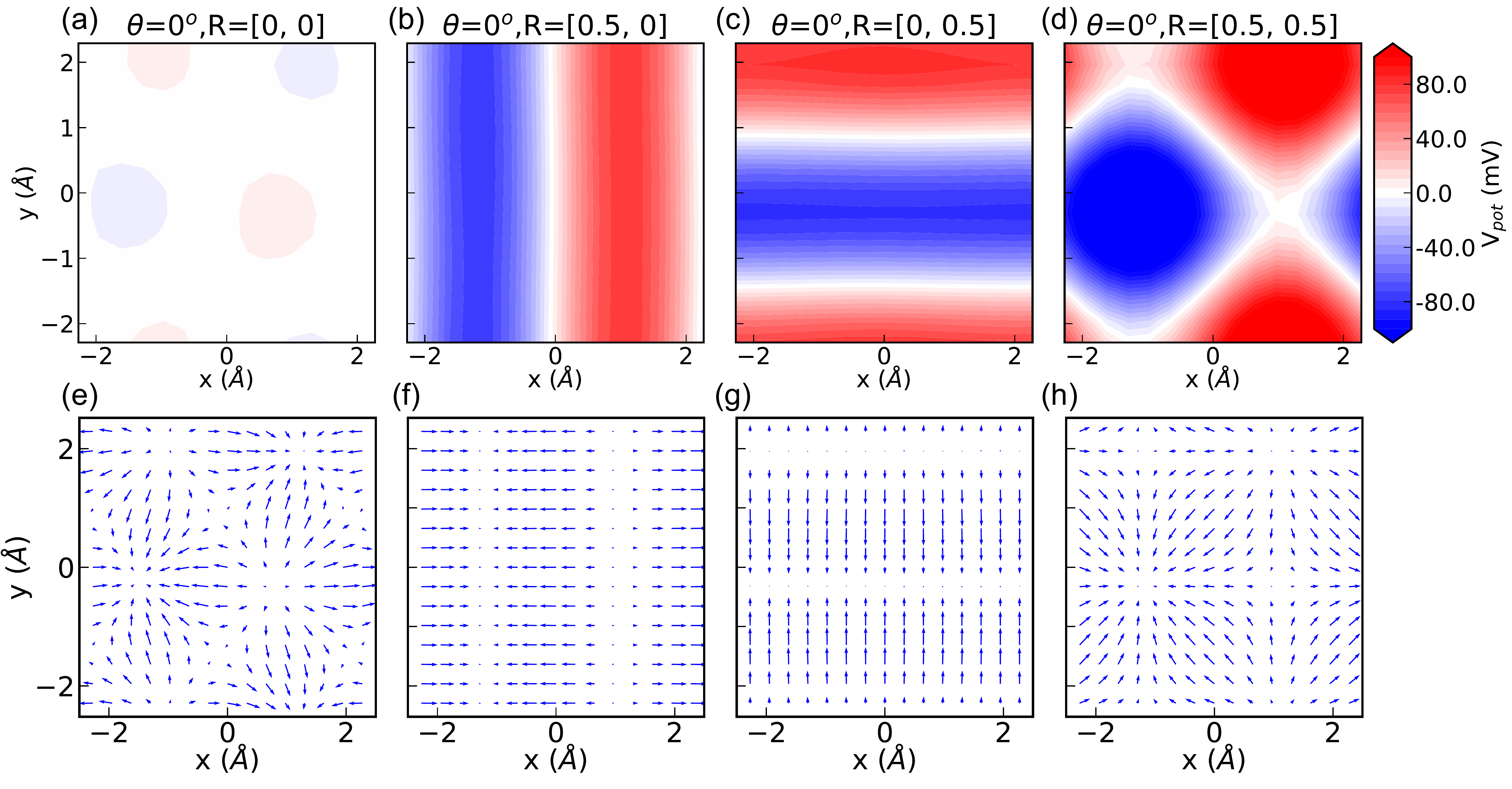}
	\caption{(a-d) Potential at the center 2D materials from two layers of SnTe $d=3.18$\AA\ above and below, see the configuration in Fig.4(a), with twist angle $\theta=0$ but different stackings where $\textbf{R}$ is the in-plane translation of the cell in fractional coordinates. (e-h) in-plane electric field where the length and direction of the arrows show the magnitude and direction of the electric field, respectively.} \label{figs:MMCt0}
\end{figure}

\begin{figure}[H]
	\centering
	\includegraphics[width=0.6\linewidth, center]{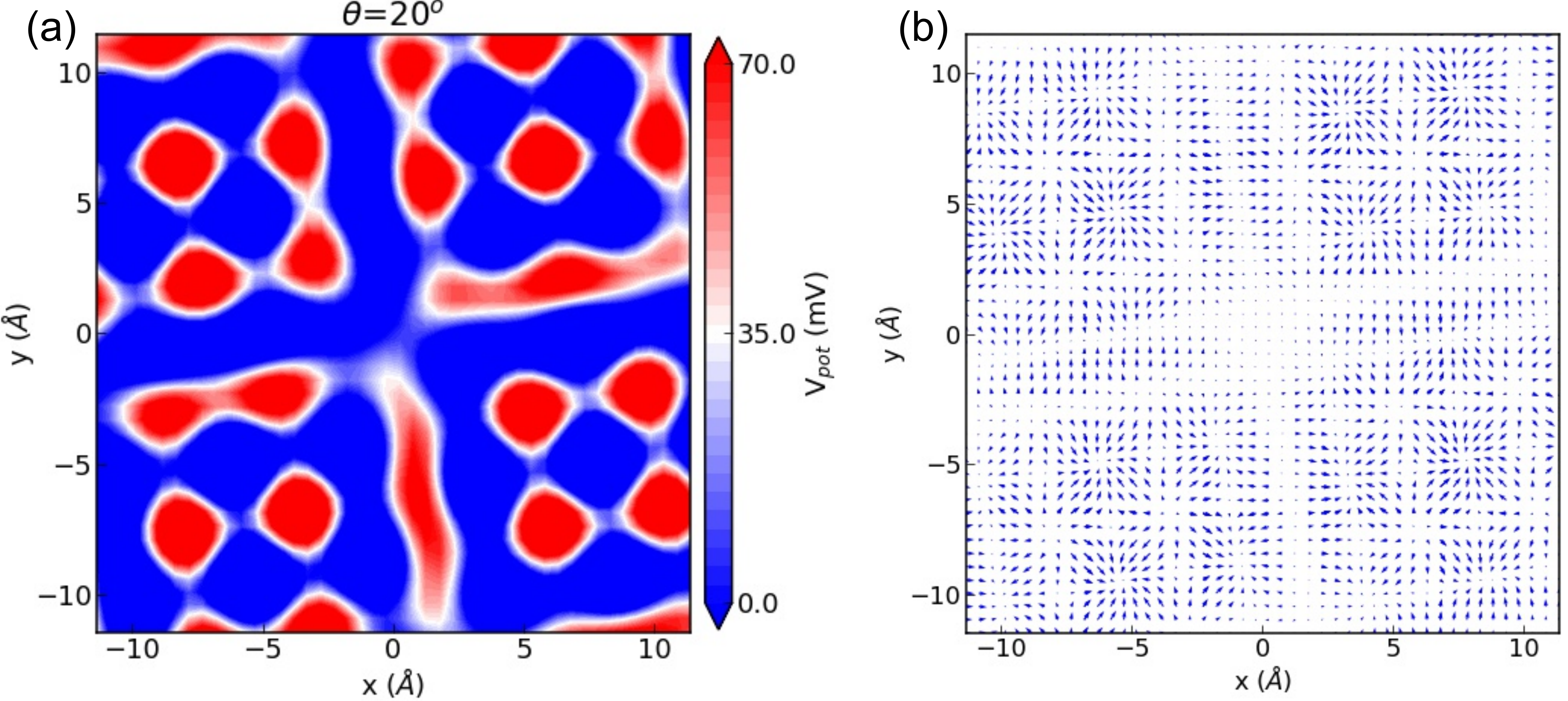}
	\caption{In-plane electric field distribution for a moir\'{e} structure. For a heterostructure as shown in Fig.4 (a) for SnTe with a twist angle $\theta=20^o$, the (a) potential and (b) in-plane electric field.} \label{figs:MMCt20}
\end{figure}

%\begin{figure}[H]
%	\centering
%	\includegraphics[width=0.65\linewidth]{Figures/FigSnX.jpg}
%	\caption{(a) top view and (b) side view of the unit cell and the positions of the point charges for metal monochalcogenides shown as black dots.} 
%%	\label{fig:MMCq}
%\end{figure}

% \newpage

%\singlespacing
\bibliography{Biblio/2DElectric.bib}